\documentclass[aps,prd,longbibliography,preprint,superscriptaddress]{revtex4-2}
\usepackage{amsmath}
\usepackage{comment}
\usepackage{siunitx}
\usepackage{graphicx}
\usepackage{hyperref}
\hypersetup{
	bookmarks=true,
	colorlinks=true,
	urlcolor=blue,
	citecolor=blue,
	linkcolor=blue, 
	pdftex,
	linktocpage=true, 
	linktoc=all,     
	hyperindex=true
}
\usepackage{upgreek}
\usepackage{romannum}
\usepackage{lipsum} 
\usepackage{braket}

\usepackage{textcomp}

\begin{document}

\title{Spin-optical dynamics and quantum efficiency of the single V1 center in silicon carbide}%
\author{Naoya Morioka}%
\thanks{These authors contributed equally to this work.}
\affiliation{Institute for Chemical Research, Kyoto University, Uji, Kyoto, 611-0011, Japan
}%
\author{Di Liu}%
\thanks{These authors contributed equally to this work.}
\affiliation{3rd Institute of Physics, IQST, and Research Center SCoPE, University of Stuttgart, 70569, Stuttgart, Germany
}%
\author{\"Oney O. Soykal}%
\affiliation{Booz Allen Hamilton, McLean, VA, 22102, USA
}%
\author{Izel Gediz}%
\affiliation{3rd Institute of Physics, IQST, and Research Center SCoPE, University of Stuttgart, 70569, Stuttgart, Germany
}%
\author{Charles Babin}%
\affiliation{3rd Institute of Physics, IQST, and Research Center SCoPE, University of Stuttgart, 70569, Stuttgart, Germany
}%
\author{Rainer St\"ohr}%
\affiliation{3rd Institute of Physics, IQST, and Research Center SCoPE, University of Stuttgart, 70569, Stuttgart, Germany
}%
\author{Takeshi Ohshima}%
\affiliation{National Institutes for Quantum Science and Technology, Takasaki, Gunma 370-1292, Japan
}%
\author{Nguyen Tien Son}%
\affiliation{Department of Physics, Chemistry and Biology, Linköping University, SE-58183, Linköping, Sweden
}%
\author{Jawad Ul-Hassan}%
\affiliation{Department of Physics, Chemistry and Biology, Linköping University, SE-58183, Linköping, Sweden
}%
\author{Florian Kaiser}%
\email{f.kaiser@pi3.uni-stuttgart.de}
\affiliation{3rd Institute of Physics, IQST, and Research Center SCoPE, University of Stuttgart, 70569, Stuttgart, Germany
}%
\author{J\"org Wrachtrup}%
\affiliation{3rd Institute of Physics, IQST, and Research Center SCoPE, University of Stuttgart, 70569, Stuttgart, Germany
}%

\date{\today}

\begin{abstract}
Color centers in silicon carbide are emerging candidates for distributed spin-based quantum applications due to the scalability of host materials and the demonstration of integration into nanophotonic resonators. Recently, silicon vacancy centers in silicon carbide have been identified as a promising system with excellent spin and optical properties. Here, we fully study the spin-optical dynamics of the single silicon vacancy center at hexagonal lattice sites, namely V1, in 4H-polytype silicon carbide. By utilizing resonant and above-resonant sub-lifetime pulsed excitation, we determine spin-dependent excited-state lifetimes and intersystem-crossing rates. Our approach to inferring the intersystem-crossing rates is based on all-optical pulsed initialization and readout scheme, and is applicable to spin-active color centers with similar dynamics models. In addition, the optical transition dipole strength and the quantum efficiency of V1 defect are evaluated based on coherent optical Rabi measurement and local-field calibration employing electric-field simulation. The measured rates well explain the results of spin-state polarization dynamics, and we further discuss the altered photoemission dynamics in resonant enhancement structures such as radiative lifetime shortening and Purcell enhancement. By providing a thorough description of V1 center’s spin-optical dynamics, our work provides deep understanding of the system which guides implementations of scalable quantum applications based on silicon vacancy centers in silicon carbide.
\end{abstract}

\maketitle

\section{Introduction}
Optically addressable solid-state spin systems are promising candidates for quantum technological applications including quantum communication, computing, and sensing \cite{Awschalom2018}. Quantum networks based on color centers rely on indistinguishable photons entangled with local spins. The entanglement generation requires precise coherent spin and optical control of the quantum spin system \cite{Kalb2017,Humphreys2018}. Another important aspect of the solid-state spins is the possibility of enhancing the spin-photon interaction by embedding the spins into the optical nanocavity. These have been demonstrated in diamond \cite{Janitz2020} and in silicon carbide \cite{Daniil2020}, but are expected to be optimized by careful design taking into account the spin-optical dynamics of color centers to determine figures of merit such as the Purcell factor and cooperativity of color centers embedded into resonators \cite{Janitz2020,Borregaard2019}. In quantum sensing applications \cite{Degen2017}, the characterization of rates is of critical interest as they determine the spin-state fluorescence contrast which governs important sensing parameters like sensitivity or experimental repetition rates \cite{Widmann2015,Neumann2009,Singh2020,Jelezko2004,Falk2013,Song2020}.

In recent years, silicon vacancy centers in silicon carbide, which is an industrially friendly and scalable material, have been found to be an emerging candidate for quantum technology applications. Single silicon vacancy centers at hexagonal sites in 4H polytype (V1) exhibit excellent spin and optical properties \cite{Nagy2019,Morioka2020} functioning at around liquid helium temperature \cite{Morioka2020,Gali2020}. However, optical coherent control and precise knowledge of internal spin-optical dynamics, which are essential for quantum applications, are not well understood for V1 centers.

This study presents coherent optical control of single V1 center by pulsed optical Rabi oscillation up to $3\pi$. In addition, we also investigated detailed spin-optical dynamics of the system including the radiative rates and spin-dependent intersystem-crossing (ISC) rates. We employed sub-lifetime laser pulses to initialize and readout the spin dynamics. In this scheme, the spin was first initialized by repeated excitation by short laser pulses at constant energy, and then two pump-probe pulses followed to infer ISC rates. As the initial population is determined only by the internal dynamics independent of the excitation strength, two laser pulses are sufficient to infer all relevant rates, which is contrary to the previously reported pump-probe method which can probe only the metastable-state lifetime \cite{Robledo2011njp,Manson2006}. In addition, with our pulse-train method, it is not necessary to resort to ad hoc assumptions of a deshelving model which tends to explain the complicated power dependence of $g^{(2)}$ function measurement \cite{Fuchs2015,Elke2012}. We further infer the quantum efficiency of V1 center based on the measured rates and local electric field at the defect position. From this, we estimate the minimum required Purcell enhancement factor for deterministic emitter-cavity coupling. Our results provide comprehensive information of internal spin-optical dynamics about V1 center and accelerate the realization of quantum technologies based on color centers in silicon carbide.

\section{Theoretical background}
\begin{center}
	\begin{figure}[h!]
		\includegraphics[trim=0 0 0 0,clip,scale=0.75]{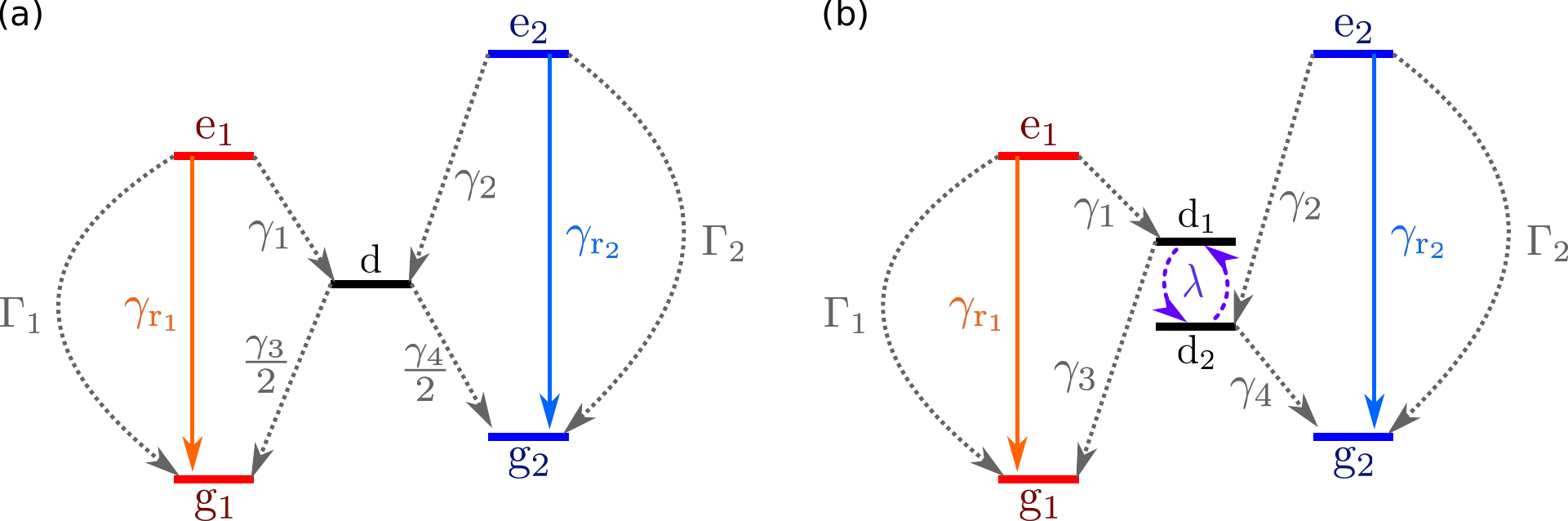}
		\caption{(a) A general five-level rate model applicable for many optically active solid-state spin defects. (b) The spin-quartet systems with $|\boldsymbol{S}|=3/2$ are further described by a six-level energy level scheme. Dashed arrows indicate all non-radiative processes while solid lines denote radiative transitions. The factor of $1/2$ for $\gamma_{3,4}$ in (a) ensures equivalence to the six-level model in (b).}
		\label{fig:theory}
	\end{figure}
\end{center}
Figure~\ref{fig:theory}(a) shows the generic five-level scheme that we use to describe the relevant dynamics for various optically-active spin systems like NV centers in diamond and divacancy centers in silicon carbide \cite{Robledo2011njp,Castelletto2015}. A ground-state qubit is provided by two long-lived spin levels (g$_{1}$, g$_{2}$). A spin-photon interface is realized through spin-conserving optical transitions to two excited states (e$_1$, e$_2$). The radiative decay rate from e$_{1,2}$ to g$_{1,2}$ is denoted by $\gamma_{\mathrm{r}_{1,2}}$. We include possible direct non-radiative decay from excited to ground states as $\Gamma_{1,2}$, e.g., based on multi-phonon relaxation \cite{McCloskey2014,Radko2016}. The intersystem-crossing (ISC) channels connect e$_1$ and e$_2$ (g$_1$ and g$_2$) levels via an effective metastable state d, with rates $\gamma_{1,2}$ ($\gamma_{3,4}$) denoting the transitions that are mediated by spin-orbit coupling and spin-spin interactions \cite{Robledo2011njp,Castelletto2015,Green2017}. 

As the V1 center is a spin-3/2 system, we assign its Kramers degenerate spin state sublevels $m_{s} = \pm1/2$ and $m_{s} = \pm 3/2$ to (g$_1$, e$_1$) and (g$_2$, e$_2$), respectively. The spin subspaces also have the same direct non-radiative decay rate, i.e., $\Gamma_{1}=\Gamma_{2}=\Gamma$, which is justified from the fact that the $A_1$ and $A_2$ transitions are energetically close and that multi-phonon decays possess no spin-selectivity (see Appendix \ref{sec:phonon}). Under group-theoretical consideration of the electronic fine structure (see Appendix \ref{sec:electronic_structure}) \cite{Nagy2019,Banks2019}, the involved metastable states are simplified into the doublets denoted as d$_1$ and d$_2$ in Fig.~\ref{fig:theory}(b). This six-level model can be further regarded as equivalent to the five-level model when the spin mixing rate $\lambda$ between the doublets is fast compared to the decay rates $\gamma_{3,4}$. For the V1 center, this equivalence between six- and five-level models will be validated later. Therefore, it is sufficient to describe the full dynamics via the radiative ($\gamma_{\mathrm{r}_{1,2}}$), the non-radiative ($\Gamma_{1,2}$) and ISC ($\gamma_{1,2,3,4}$) rates based on the five-level model.

\section{Experimental setup}
\begin{center}
	\begin{figure}[h!]
		\includegraphics[trim=0 0 0 0,clip,scale=0.07]{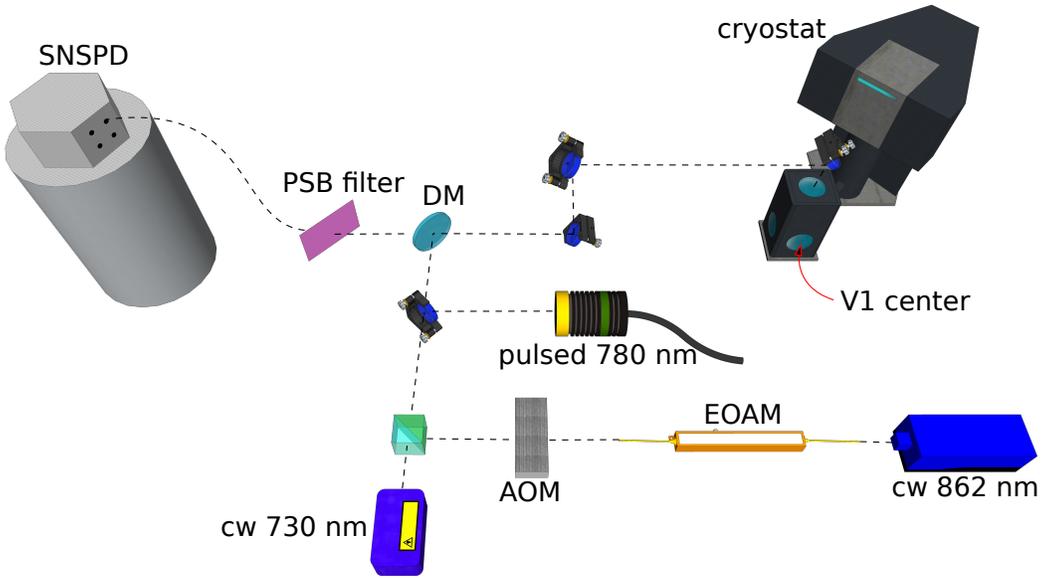}
		\caption{The confocal setup with main instruments sketched in this work (for details see the main text).}
		\label{fig:setup}
	\end{figure}
\end{center}
For our spin-optical investigations, we use a home-built confocal microscopy setup. A slightly n-type 4H-SiC sample is electron irradiated to contain a low density of individually addressable V1 centers \cite{Nagy2019,Morioka2020}. The sample is cooled down to a temperature of $T=\SI{5}{K}$ in a closed-cycle cryostat (Montana Instruments). Due to the V1 center’s relatively small ground-state zero field splitting of $\SI{4.5}{MHz}$, we apply a small external magnetic field of about 9 Gauss along the crystal $c$-axis to avoid undesired hybridization caused by Earth’s magnetic field \cite{Nagy2019,Morioka2020}. To improve light collection efficiency, we fabricate microscopic solid immersion lenses (SILs) on top of V1 centers \cite{Siyushev2010}. 

Figure~\ref{fig:setup} shows the experimental setup toward inferring the involved rates using three key laser excitation systems: (i) A $\SI{730}{nm}$ continuous-wave laser is used for prolonged above-resonant excitation, which has been identified as a suitable method to depolarize the V1 center’s ground state \cite{Nagy2019}. (ii) A $\SI{780}{nm}$ picosecond pulsed laser (PicoQuant LDH-P-C-780, $\SI{56}{ps}$ intensity full width at half maximum (FWHM)) is used for initial state preparation and evaluating ISC rates via pulse-train measurements, which we introduce in this study. (iii) Spin-selective resonant laser pulses at $\SI{862}{nm}$ with $\SI{1.5}{ns}$ Gaussian intensity FWHM are used to infer radiative decay rates and the transition dipole moment. The pulses are obtained by shaping a cw resonant laser (Toptica DL pro) with an electro-optic amplitude modulator (EOAM, Jenoptik). The pulse amplitude is controlled using an acousto-optic modulator (AOM, Gooch \& Housego). For all measurements, fluorescence emission is detected in the red-shifted phonon sidebands (PSBs) $880-\SI{1000}{nm}$ \cite{Gali2020} using long-pass filters and a superconducting nanowire single-photon detector (SNSPD, Photon Spot).

\section{Experimental results}
\subsection{Spin-selective relaxation out of the metastable state}	
A previous study demonstrated that prolonged above-resonant excitation eventually populates the ground states g$_1$ and g$_2$ of V1 equally \cite{Nagy2019}. This observation is explained by a metastable-state (MS) lifetime that is significantly longer than the excited-state (e$_1$, e$_2$) lifetimes, such that the system is mainly populated in the MS during excitation. After the excitation, the system relaxes into the ground states according to the ratio $\gamma_{3}/\gamma_{4}$ of ISC rates out of the MS. The observed spin depolarization implies therefore that:
\begin{equation}\label{eq:gamma3gamma4}
	\gamma_{3} \simeq \gamma_{4},
\end{equation}	
which is also supported by theory (see Appendix \ref{sec:electronic_structure}).
\subsection{Spin-dependent excited-state lifetimes}
In the next step, we investigate the excited-state lifetimes $\tau_\mathrm{e_{1,2}}=(\gamma_{\mathrm{r}_{1,2}}+\Gamma_{1,2}+\gamma_{1,2})^{-1}$. Previously, above-resonant short pulses were used which only gave a spin-nonselective lifetime of $\SI{5.5}{ns}$ \cite{Nagy2018}. Here, we provide a much deeper insight of the system by probing spin-selective lifetimes. Fig.~\ref{fig:lifetime}(a) shows the lifetime-measurement sequence. We initially depolarize the ground states by $\SI{40}{\upmu s}$-long above-resonant excitation at $\SI{730}{nm}$, followed by a spin-selective resonant excitation pulse between either g$_1$ and e$_1$ ($A_1$ transition) or g$_2$ and e$_2$ ($A_2$ transition). From the time-dependent fluorescence decay shown in Fig.~\ref{fig:lifetime}(b), we infer the spin-dependent lifetimes to be
\begin{equation*}\label{eq:lifetime1}
	\tau_\mathrm{e_{1}} = (\gamma_{\mathrm{r}_{1}}+\Gamma_{1}+\gamma_{1})^{-1} = \SI{5.03\pm0.02}{ns}
\end{equation*}
and
\begin{equation*}\label{eq:lifetime2}
	\tau_\mathrm{e_{2}} = (\gamma_{\mathrm{r}_{2}}+\Gamma_{2}+\gamma_{2})^{-1} = \SI{6.26\pm0.02}{ns},
\end{equation*}
respectively by single exponential fits. The shorter excited-state lifetime $\tau_\mathrm{e_{1}}$ explains the commonly observed larger linewidth for the $A_{1}$ optical transition \cite{Nagy2019,Morioka2020}. 

\subsection{Optical Rabi measurement to probe spin-dependent radiative decay rates}
\begin{center}
	\begin{figure}[h!]
		\includegraphics[trim=0 0 0 0,clip,scale=0.75]{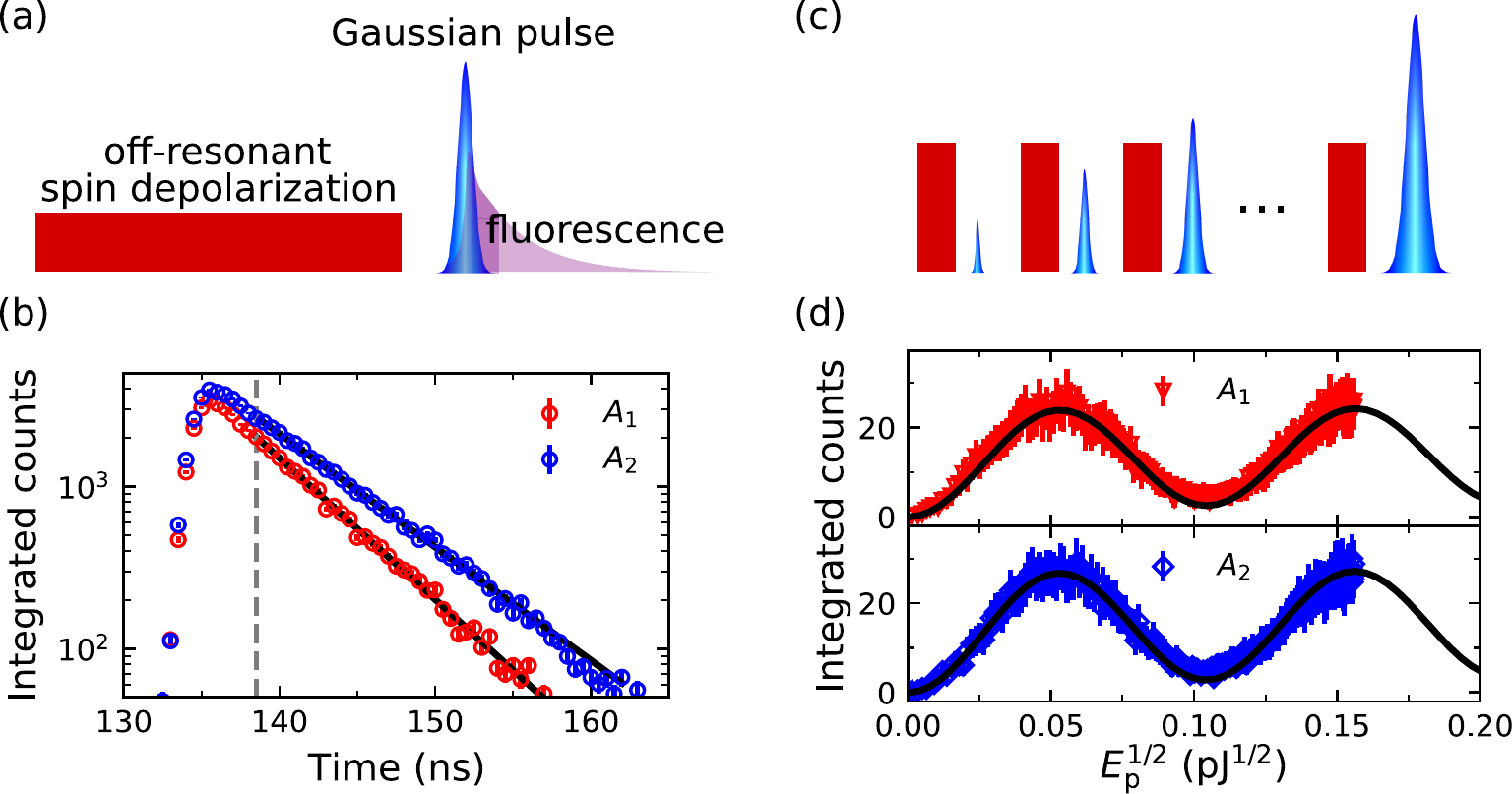}
		\caption{Excited-state lifetime and optical Rabi measurements. (a) Measurement sequence for the spin-selective excited-state lifetimes. (b) Fluorescence emission from excited states with single exponential fits. (c) Measurement sequence for optical Rabi oscillations. The ground-state spin is initialized by the above-resonant laser. (d) Fluorescence emission as a function of the square root of the calibrated pulse energy ($E_\mathrm{p}^{1/2}$). Black curves are from simulation based on quantum master equations \cite{Banks2019}.}
		\label{fig:lifetime}
	\end{figure}
\end{center}

The next measurement is outlined to evaluate the spin dependence of the radiative decay rates $\gamma_{\mathrm{r}_{1,2}}$. To this end, we conduct pulsed optical Rabi measurement as a function of the laser pulse energy. As shown in Fig.~\ref{fig:lifetime}(c), each sequence starts by depolarizing the ground states with $\SI{730}{nm}$ above-resonant excitation, followed by a $\SI{1.5}{ns}$ Gaussian resonant pulse whose amplitude is controlled by an AOM. Fig.~\ref{fig:lifetime}(d) shows the integrated fluorescence emission as a function of the square root of the pulse energy. Note that we gate the fluorescence detection at $\SI{3}{ns}$ after the peak of the excitation pulse to avoid signal deterioration due to laser breakthrough. The results show clear coherent optical Rabi oscillations up to $3\pi$. Rabi frequencies are identical for the two spin manifolds and the $\pi$-pulse energy is $\SI{2.8\pm0.1}{fJ}$. This implies that both transitions have the same transition dipole strength, or in other words, the radiative decay rates are spin independent: $\gamma_{\mathrm{r}_{1}} = \gamma_{\mathrm{r}_{2}}$ which in agreement with theory (see Appendix \ref{sec:appendix4}). In the discussion section \ref{sec:discussion}, we use these results to further infer the V1 center’s transition dipole moment.

\subsection{Pulse-train scheme for deterministic spin initialization and inferring ISC rates and MS lifetime}
\begin{center}
	\begin{figure}[h!]
		\includegraphics[trim=0 0 0 0,clip,scale=0.75]{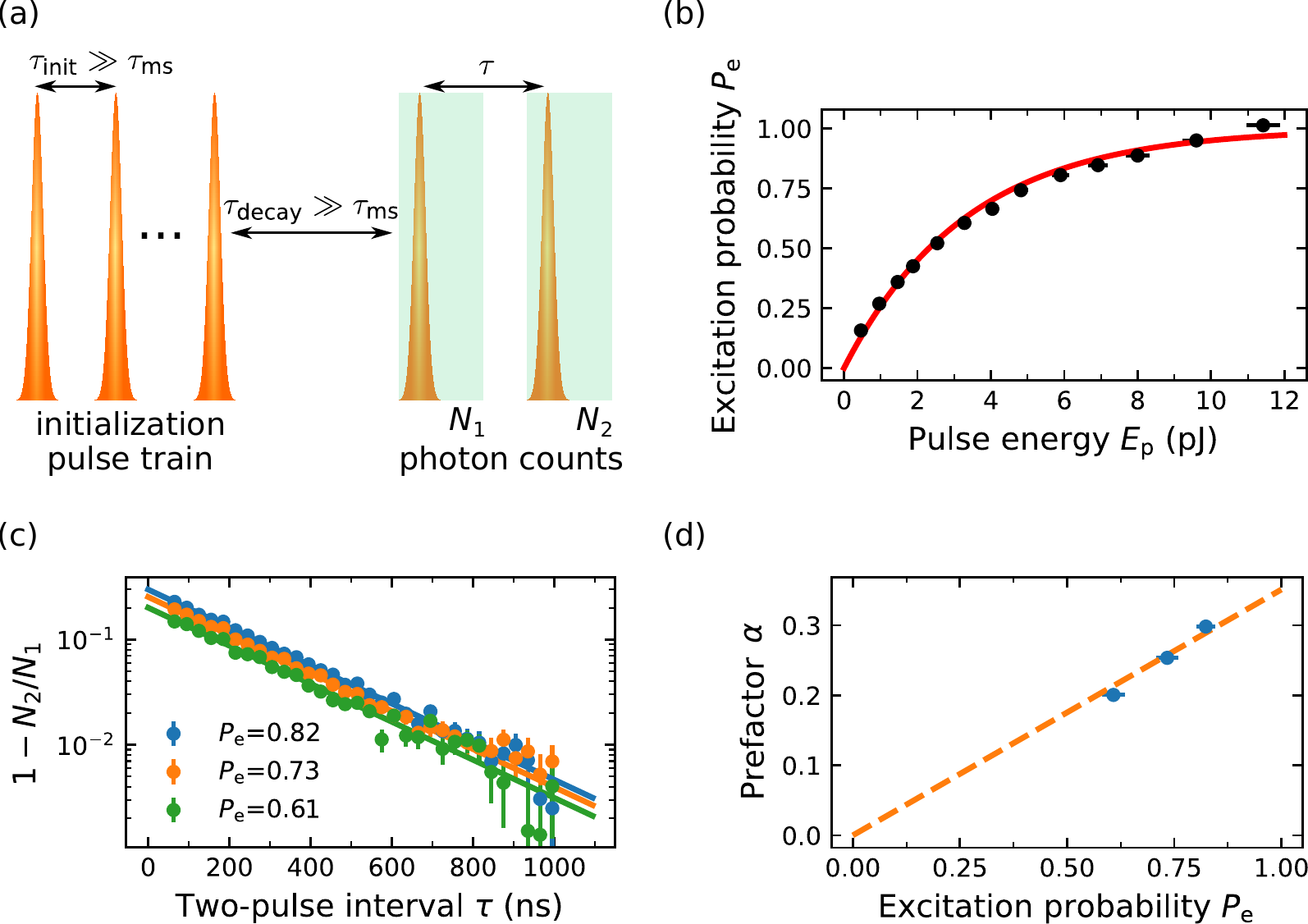}
		\caption{Deterministic spin initialization with pulse-train scheme and pump-probe measurement results. (a) The pulse sequence of the measurement. The system is initialized by pulses with long intervals and relaxed into the ground states. Then, the system is excited with two consecutive pulses for 'pump' and 'probe' to extract the rates. (b) The pulsed excitation probability with its saturation fit. (c) The fluorescence ratio data $1-N_{2}/N_{1}$ are shown for three different excitation probabilities after the background subtraction and pulse energy correction (see Appendix \ref{sec:appendix4}) with their single exponential fits. (d) The prefactor of the exponential decay $\alpha$ obtained from fitting in panel (c) depicts its linear dependence of the excitation probability with the slope determined by the rates.}
		\label{fig:pulsetrain}
	\end{figure}
\end{center}
The above-described measurements are not sufficient to infer the individual ISC rates. In order to do so, now we introduce a pulse-train measurement consisting of sub-lifetime short pulses. The underlying idea of our approach is based on the following two steps: 1. First, a pulse train is used to deterministically initialize the spin ground states into a steady-state population determined only by the system’s internal transition rates (see Appendix \ref{sec:theory_pulsetrain}). 2. Thereafter, the system is excited (‘pump’) by the first measurement pulse (same as those used for initialization). The decay back to the ground state with a decay rate dominated by the MS lifetime is read out (‘probe’) by the second measurement pulse. Our scheme may appear similar to previous pump-probe experiments with two sub-microsecond long pulses \cite{Robledo2011njp,Manson2006}. In contrast, our scheme can also determine the branching ratio of the ISC rates from the excited states by employing deterministic spin initialization with sub-lifetime short pulses. Together with the lifetime measurement, we can determine the spin-selective ISC rates. Note that our scheme is different from a pulse-train technique used for molecule ensembles \cite{GOTARDO2017}. 

Next, we introduce our pulse-train and pump-probe measurement in detail. For initialization of the ground state with a pulse train, we use multiple \SI{56}{ps} short pulses, which are significantly shorter than the excited state lifetimes $\tau_\mathrm{e_{1,2}}$. To ensure that the system resides in the ground states whenever a pulse arrives, we fix the pulse repetition interval at 1µs, which is significantly longer than the MS lifetime $\tau_\mathrm{ms}$. Importantly, this avoids complicated dynamics such as re-excitation from the metastable state to other higher lying states \cite{Fuchs2015}. As aforementioned, the key advantage of this initialization method is that the steady-state ground-state spin populations are determined solely by the internal dynamics of the spin system, regardless of the excitation power. The used optical pulse power range in our measurement does not saturate the optical transition and clear single-exponential behavior is observed in the saturation measurement (see below), which implies a negligible ionization effect requiring two-photon absorption \cite{Niethammer2019}. In addition, this initialization method is not affected by the existence of optical deshelving processes from the metastable state, which were proposed to explain the discrepancies in the $g^{(2)}(\tau)$ data for silicon vacancies in 4H-SiC at cubic lattice sites (V2) \cite{Fuchs2015}. Therefore, our method contributes to accurate and unambiguous extraction of the involved rates. Theoretical spin initialization dynamics is analytically discussed in Appendix \ref{sec:theory_pulsetrain}.

After the initialization, the two consecutive $\SI{56}{ps}$-short measurement pulses excite the system at an inter-pulse delay $\SI{65}{ns}<\tau<\SI{1000}{ns}$. The first pulse is used to read out the steady-state ground-state population after the pulse train, and the second pulse infers the time-dependent decay from the metastable state to the ground state through fluorescence. The relative fluorescence intensity from both pulses ($N_{2}/N_{1}$) is then given by a single exponential function for $\tau \gg \tau_\mathrm{e_{1,2}}$:
\begin{equation}\label{eq:ratioN1N2}
	1-\frac{N_{2}}{N_{1}} = \alpha e^{-\tau/\tau_\mathrm{ms}}.
\end{equation}	
Here, the decay constant is the MS lifetime $\tau_\mathrm{ms}^{-1}=(\gamma_{3}+\gamma_{4})/2$. Additionally, with the initial state prepared by the pulse-train method, the prefactor $\alpha$ allows to also extract the intersystem-crossing rates $\gamma_{1}$ and $\gamma_{2}$ through the relation:
\begin{equation}\label{eq:alpha}
	\alpha = P_\mathrm{e} \frac{\gamma_{3}\tau_\mathrm{e_{1}} + \gamma_{4}\tau_\mathrm{e_{2}}}{\gamma_{3}/\gamma_{1} + \gamma_{4}/\gamma_{2}} \frac{1- (\gamma_{3}\tau_\mathrm{e_{2}} +\gamma_{4}\tau_\mathrm{e_{1}})/2}{ (1-\tau_\mathrm{e_{1}}/\tau_\mathrm{ms} ) (1-\tau_\mathrm{e_{2}}/\tau_\mathrm{ms} ) }.
\end{equation}
Here, the excitation probability per pulse is $P_\mathrm{e}$, which is determined from an independent fluorescence saturation measurement at a slow repetition rate of $\SI{0.5}{MHz}$ to ensure that the system resides in the ground state at each excitation. As shown in Fig.~\ref{fig:pulsetrain}(b), the fluorescence rate saturates exponentially as $\propto(1-e^{-E_\mathrm{p}/E_{0}})$, indicating negligible ionization behavior. Again, in this readout part, potential deshelving processes do not affect the result because the metastable state population is empty at the first pulse and no optical excitation occurs between the two pulses.

To underline the validity of our method, we perform the two-pulse excitation measurement at three different excitation probabilities $P_\mathrm{e}$. The results are presented in Fig.~\ref{fig:pulsetrain}(c). All measurements can be fitted with the same exponential decay time, i.e., the MS lifetime is independent of the pulse power. Moreover, the data in Fig.~\ref{fig:pulsetrain}(d) corroborate that the prefactor $\alpha$ shows the predicted linear dependence on the excitation probability in Eq.~\ref{eq:alpha}.

With $\gamma_{3}\simeq\gamma_{4}$ (Eq.~\ref{eq:gamma3gamma4}) for V1 center in 4H-SiC known from previous work \cite{Nagy2019} and validated by theory (see Appendix \ref{sec:electronic_structure}), all the ISC rates can now be determined. We note that our approach can also be applied for systems that show $\gamma_{3}\neq\gamma_{4}$, e.g., by using resonant laser pulses to differentiate the spin-dependent populations (see Appendix \ref{sec:appendix5}). A summary of the results obtained for single V1 center in 4H-SiC is given in Tab.~\ref{tab:rates}.

\begin{table}
	\centering
	\caption{Summary of radiative, non-radiative and ISC rates and MS lifetime for single V1 center in 4H-SiC.}
	\label{tab:rates}
	\begin{tabular}{cc}
		\hline\hline
			  & Experimental (ns)\\\hline
		e$_{1,2} \rightarrow$ g$_{1,2}$: $(\gamma_\mathrm{r}+\Gamma)^{-1}$ & $9.0\pm0.1$  \\ 
		e$_{1} \rightarrow$ d: $\gamma_{1}^{-1}$ & $11.4\pm0.2$  \\
		e$_{2} \rightarrow$ d: $\gamma_{2}^{-1}$ & $20.5\pm0.6$  \\
		$\tau_\mathrm{ms}$ & $240\pm2$\\
		\hline\hline  
	\end{tabular}
\end{table}

\subsection{Resonant depletion and ground-state initialization measurements}
\begin{center}
	\begin{figure}[h!]
		\includegraphics[trim=0 0 0 0,clip,scale=0.75]{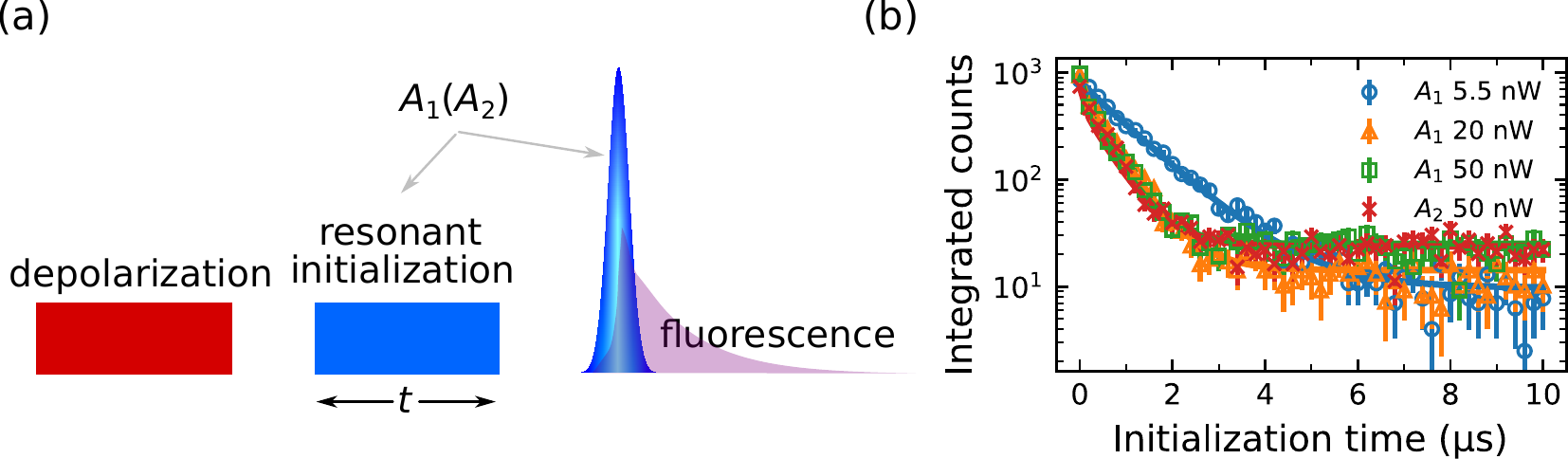}
		\caption{(a) Resonant depletion measurement sequence (see the main text for detailed explanation). (b) The remaining ground-state population after depletion into the dark state as a function of the initialization time at different initialization powers. The solid lines denote the simulated curves using the experimental obtained rates in Tab.~\ref{tab:rates}.}
		\label{fig:depletion}
	\end{figure}
\end{center}
With the above obtained rates, we gain an improved understanding of the photoemission and spin-optical dynamics of single V1 center in silicon carbide. Next, we present near-deterministic ground-state spin initialization scheme based on spin-selective resonant depletion and validate the equivalence between the six- and five-level rate model employed for the defect system. As shown in the six-level scheme in Fig.~\ref{fig:theory}(b), the optical spin-pumping is mediated via ISC through the metastable states. The time constant for initialization into the ground spin ($m_{s}\pm1/2$ or $m_{s}\pm3/2$) depends directly on the spin mixing rate $\lambda$ between the doublets d$_{1}$ and d$_{2}$. As mentioned before, the five-level model in Fig.~\ref{fig:theory}(a) is justified when $\lambda\gg\gamma_{3,4}$.

Our resonant depletion measurement sequence is shown in Fig.~\ref{fig:depletion}(a). First, a prolonged above-resonant (\SI{730}{nm}) laser pulse depolarizes the ground-state spins. Second, we perform optical spin-pumping using quasi-cw resonant excitation at different power levels for a time duration $0\leq \tau_\mathrm{init} \leq \SI{10}{\upmu s}$ on the $A_{1}$ ($A_{2}$) transition (see Appendix \ref{sec:appendix6}). Via ISC, this eventually increases the spin population in the opposite ground state g$_{2}$ (g$_{1}$). After the system completely decays from the metastable-state manifold, we read out the ground-state spin population with a $\SI{1.5}{ns}$ resonant laser pulse on the $A_{1}$ ($A_{2}$) transition. Compared to the previous work \cite{Banks2019}, where a cw resonant laser serves for optical spin-pumping and readout simultaneously, our depletion sequence avoids complicated dynamics by using sub-lifetime short resonant pulses to read out the ground-state population more accurately. The readout laser pulse energy was kept constant regardless of the quasi-cw laser power during the spin-pumping procedure. 

Experimental results for $A_{1}$ excitation at three different power levels is shown in Fig.~\ref{fig:depletion}(b). All three data sets show the expected decay of the readout fluorescence as a function of the initialization time. At longer times, the initialization fidelity into the ground state g$_2$ reaches $99.2\pm0.3\%$, which is comparable to the initialization fidelity in NV \cite{Robledo2011}. The small degradation of the initialization fidelity at higher powers can be explained by optical power broadening leading to reduced spin selectivity during optical excitation. Focusing on the power-dependency of the spin initialization time, we find that at \SI{5.5}{nW} the time constant is dominated by the optical excitation rate. Conversely, for pump powers larger than $\SI{20}{nW}$, no further shortening of the initialization time is observed as system dynamics are now dominated by the excited-state and metastable-state lifetimes. We calculate the depletion time evolution by solving a quantum master equation of the six-level model with the rates obtained by the pulse-train measurements and setting a fast spin mixing rate \cite{Banks2019}. The results for all the initialization powers, presented in Fig.~\ref{fig:depletion}(b) as solid lines, show excellent agreement with the experimental data. This result confirms that the obtained rates by the pulse-train method are reliable and that the mixing rate between the doublets is significantly faster than the decay rates $\gamma_{3,4}$, such that the equivalent five-level model is justified.

Alternatively, we can obtain the rates by utilizing the experimental resonant depletion data at different powers to calculate the density matrix master equation using a custom Nelder-Mead and differential evolution optimization algorithm. With the fine parameter fitting, similar rates were obtained as those measured by the pulse-train measurement scheme (see Appendix \ref{sec:appendix7}).

\section{Discussion\label{sec:discussion}}
In this section we discuss how to employ the measured rates to estimate the quantum efficiency of the V1 center, which in turn gives the minimum required Purcell enhancement factor for deterministic emitter-cavity coupling \cite{Janitz2020,Sipahigil2016,Bracher2017,Crook2020,Lukin2020}. The quantum efficiency (QE) of an emitter is defined as the ratio between the radiative and the sum of all non-radiative processes (including ISC) from the excited states (e$_{1,2}$ in the case of V1) \cite{Janitz2020}. Notably, nonradiative processes can include direct phonon relaxation \cite{Radko2016}, which is generally not straightforward to measure and may explain the large deviation in the estimated quantum efficiencies of nitrogen vacancy centers \cite{Mohtashami2013,INAM2014} and silicon vacancy centers \cite{Elke2012,Alkauskas_2014} in diamond. Our estimation gives the radiative and nonradiative lifetimes as follows.

Firstly, we estimate the transition dipole moment from the Rabi oscillation. For this purpose, precise knowledge on the local time-dependent electrical field strength $E(t)$ at the defect position is necessary. To estimate the field in our SIL geometry, we first use finite-difference frequency domain (FDFD) simulations to obtain the field strength $E_\mathrm{bulk}$ for defects in the bulk without SIL. By comparing optical saturation powers of defects in the bulk and in a SIL, we obtain the field in the SIL as $E_\mathrm{SIL} = E_\mathrm{bulk}\sqrt{P_\mathrm{sat,bulk}/P_\mathrm{sat,SIL}} \leq \SI{8.8}{kV\cdot m^{-1}}$ (see Appendix \ref{sec:appendix8}). Moreover, we consider that the local effective field at the defect is different from the applied field due to the polarizability of lattices \cite{Stoneham2007}. To estimate the local effective field at the defect, we use the Lorentz-Lorenz model \cite{Stoneham2007}, i.e. $E_\mathrm{local}/E_\mathrm{SIL} = (\epsilon+2)/3$, to estimate the field correction factor as 2.92 leading to the local effective field $E_\mathrm{local}=\SI{26}{kV\cdot m^{-1}}$. Within the area theorem and under the rotating wave approximation \cite{Fischer2017}, the transition dipole moment is given by 
\begin{equation}\label{eq:dipole_moment}
	\mu = \frac{\pi \hbar }{\int E_\mathrm{local}\cdot e^{-\frac{t^{2}}{2\sigma_{E}^{2}}} \mathrm{d}t}
\end{equation}
where $\sigma_{E} = \omega_\mathrm{pulse}/(2\sqrt{\ln 2})$ and $\omega_\mathrm{pulse} = \SI{1.5}{ns}$. Therefore, the lower bound of the zero-phonon-line transition dipole moment is $\mu_\mathrm{ZPL_{1,2}}^\mathrm{lower} = \SI{0.36}{\AA}$. With the relation of the dipole moment and the spontaneous emission rate $\gamma_\mathrm{ZPL} = (n\omega^{3} \mu_\mathrm{ZPL}^{2})/( 3\epsilon_{0} \pi c^{3} \hbar)$ \cite{Alkauskas_2014}, the ZPL radiative decay rate is calculated to be $\gamma_\mathrm{ZPL_{1,2}} \geq (\SI{270}{ns})^{-1}$. With the previously measured V1 center’s Debye Waller factor (DWF) of 8\% \cite{Gali2020}, the total radiative decay rate for both transitions is $\gamma_{\mathrm{r}_{1,2}}= \gamma_\mathrm{r} = \gamma_\mathrm{ZPL} /\mathrm{DWF} \geq (\SI{21}{ns})^{-1}$. Thus, the total transition dipole moment of V1 is $\mu _\mathrm{total} \geq \SI{1.3}{\textit{e}\AA}$, which is comparable to the transition dipole moment of NV ($\sim\SI{1.08}{\textit{e}\AA}$) in diamond \cite{Alkauskas_2014}.

Next, we show that the measured rates can be used to estimate the system’s quantum efficiency and to set an upper bound on the rate of all other relaxation processes from the two excited spin subspaces $\Gamma_{1,2}=\Gamma$ as indicated in the five-level scheme shown in Fig.~\ref{fig:theory}b. With $\gamma_\mathrm{r}\geq (\SI{21}{ns})^{-1}$ and the sum of the radiative and direct non-radiative decay rates obtained from the pulse-train measurement $\gamma_\mathrm{r} +\Gamma = (\SI{9.0}{ns})^{-1}$, the additional direct non-radiative decay rate is constrained to be $\Gamma \leq (\SI{16}{ns})^{-1}$. Since the definition of the internal quantum efficiency (QE) of defects is given by the ratio of the radiative decay rate to the total decay rate, the lower bounds of QE corresponding to both transitions are:
\begin{align*}\label{eq:QEs}
	\mathrm{QE_{1}} &= \gamma_\mathrm{r} \cdot  \tau_\mathrm{e_{1}} \geq 23\%,\\
	\mathrm{QE_{2}} &= \gamma_\mathrm{r} \cdot  \tau_\mathrm{e_{2}} \geq 29\%.
\end{align*}
The quantum efficiency of V1 center in silicon carbide is thus similar to that of silicon vacancy centers in diamond (29.6\%) \cite{Becher2017}, which promises great potential for integration into photonic crystal cavities \cite{Sipahigil2016}.

Furthermore, we also estimate the overall collection efficiency of our confocal setup using the total radiative decay rate $\gamma_\mathrm{r} \geq (\SI{21}{ns})^{-1}$ derived from the local-field calibration. The PSB saturation photoemission rate under cw illumination is expected to be $I_\mathrm{PSB}^{\infty} = (1-\mathrm{DWF}) \cdot \eta_\mathrm{det} \cdot \gamma_\mathrm{r} \cdot (n_\mathrm{e_1}^{\infty} + n_\mathrm{e_2}^{\infty}) \geq \eta_\mathrm{det} \cdot \SI{2.7}{MHz}$ with the steady-state excited-state populations $n_\mathrm{e_1}^{\infty} = 2.2\%$ and $n_\mathrm{e_2}^{\infty} = 4.0\%$ (simulated from the measured rates). After optimizing the alignment of the setup, the saturation photoemission detection rate was measured $I_\mathrm{PSB}^{\infty} = \SI{33}{kHz}$. Thus, the overall detection efficiency of our setup is $\leq 1.2\%$. This upper bound is comparable to state-of-the-art work using NVs \cite{Humphreys2018}, in which the overall PSB photon collection efficiency is about $1.5\%$ (calculated from the ZPL photon collection efficiency $4\times10^{-4}$ \cite{Humphreys2018}, and the DWF of NV centers 2.55\% \cite{Riedel2017}).

From the obtained rates and previously determined pure dephasing rate \cite{Morioka2020} and the Debye-Waller factor \cite{Gali2020} of V1 center, we evaluate the minimum Purcell factor required to realize efficient coupling between the emitter and a cavity which enhances ZPL. To realize the deterministic coupling, i.e. cooperativity larger than 1, Purcell factors of 54 and 43 are required for $A_1$ and $A_2$ ZPL transitions, respectively. The shortened excited-state lifetimes are $\SI{2.5}{ns}$ and $\SI{3.2}{ns}$, respectively, which still permit spin-selective excitation. The prior knowledge of the minimum required Purcell factor directly guides the design of V1 center integrated photonic crystal cavities in silicon carbide as the Purcell factor is proportional to the ratio of the quality factor ($Q$) to the mode volume ($V$) of the cavity.

We also estimated the non-radiative rate $\Gamma$ by using equations derived for the multiphonon relaxation process to be in the order of ms (see Appendix \ref{sec:phonon}), which deviates from the upper boundary value estimated from the Rabi measurement. This discrepancy may be attributed to the overestimation in the local electric field due to the Lorentz-Lorenz correction \cite{Stoneham2007} or additional field shielding around the V1 center considering that the dominant electron traps of the sample are carbon vacancies \cite{Nagy2019,Morioka2020}. Another explanation for the measured sizeable value of $\Gamma$ may involve additional decay mechanisms that could be investigated in future work. 

\section{Conclusion}
In this work we studied the spin-optical dynamics of single V1 center in silicon carbide in detail based on the five-level rate model which is equivalent to an energy level scheme derived from group theory for spin-3/2 color centers. The excited-state lifetimes were measured in a spin-selective manner with sub-lifetime short resonant pulses and the values well explain the difference in the optical linewidths of the spin subspaces 1/2 and 3/2. In order to infer the intersystem-crossing rates and the metastable-state lifetime, we developed a pulse-train scheme employing sub-lifetime short pulses, which allowed us to measure the spin-selective spin-optical rates while avoiding complicated re-excitation into higher-lying states as opposed to measuring with other existing techniques.

Based on the measured rates, we also estimated the optical transition dipole moment and gave a lower bound of the quantum efficiency of single V1 center. We further gave an estimation on the minimum Purcell enhancement factor, providing a guideline for the optimal nanophotonic cavity design embedding V1 color center in silicon carbide. In addition, we demonstrated coherent optical control and near-unity spin-subspace initialization of V1 center, which proves the robustness of the system for quantum applications. Our work also paves the way for studying photophysics and the intersystem-crossing mechanism of existing and emerging defects with high accuracy and is of great importance for quantum technologies based on solid-state spin-active emitters and quantum communication.

\section*{acknowledgements}
We thank Sreehari Jayaram, Vladislav Bushmakin, Jonathan K\"orber, Matthias Niethammer, Dr. Matthias Widmann, Dr. Jianpei Geng, Dr. Vadim Vorobyov and Prof. Dr.-Ing. Roland Nagy for helpful discussion. D.L., I.G., C.B., R.S., F.K. and J.W. acknowledge the EU-FET Flagship on Quantum Technologies through the project ASTERIQS (Grant Agreement No. 820394), the European Research Council (ERC) grant SMel, the Max Planck Society, and the German Research Foundation (SPP 1601, FOR 2724). F.K. and J.W. acknowledge support by the EU-FET Flagship on Quantum Technologies through the project QIA (Grant Agreement No. 820445), the Baden-W\"urttemberg Foundation for the projects QT-6: SPOC and QC4BW (Grant Agreement No. 3-4332.62-IAF/7), as well as the German Federal Ministry of Education and Research (BMBF) for the projects QR.X (Grant Agreement No. 16KISQ008), Spinning (Grant Agreement No. 13N16219) and QVOL (Grant Agreement No. 03ZU1110IB). J.U.H. acknowledges the Swedish Energy Agency (Grant No. 43611-1) and Swedish Research Council (Grant No. VR 2020-05444). N.T.S. acknowledges the Swedish Research Council (Grant No. VR 2016-04068). N.T.S. and J.U.H. thank the EU H2020 FETOPEN project QuanTELCO (Grant No. 862721) and the Knut and Alice Wallenberg Foundation (Grant No. KAW 2018.0071). T.O. acknowledges support by JSPS KAKENHI (Grant No. 20H00355).

\begin{appendix} 		
\section{Full electronic structure of the V1 defect in 4H-SiC\label{sec:electronic_structure}}
\begin{center}
\begin{figure}[h!]
		\includegraphics[trim=0 0 0 0,clip,scale=0.75]{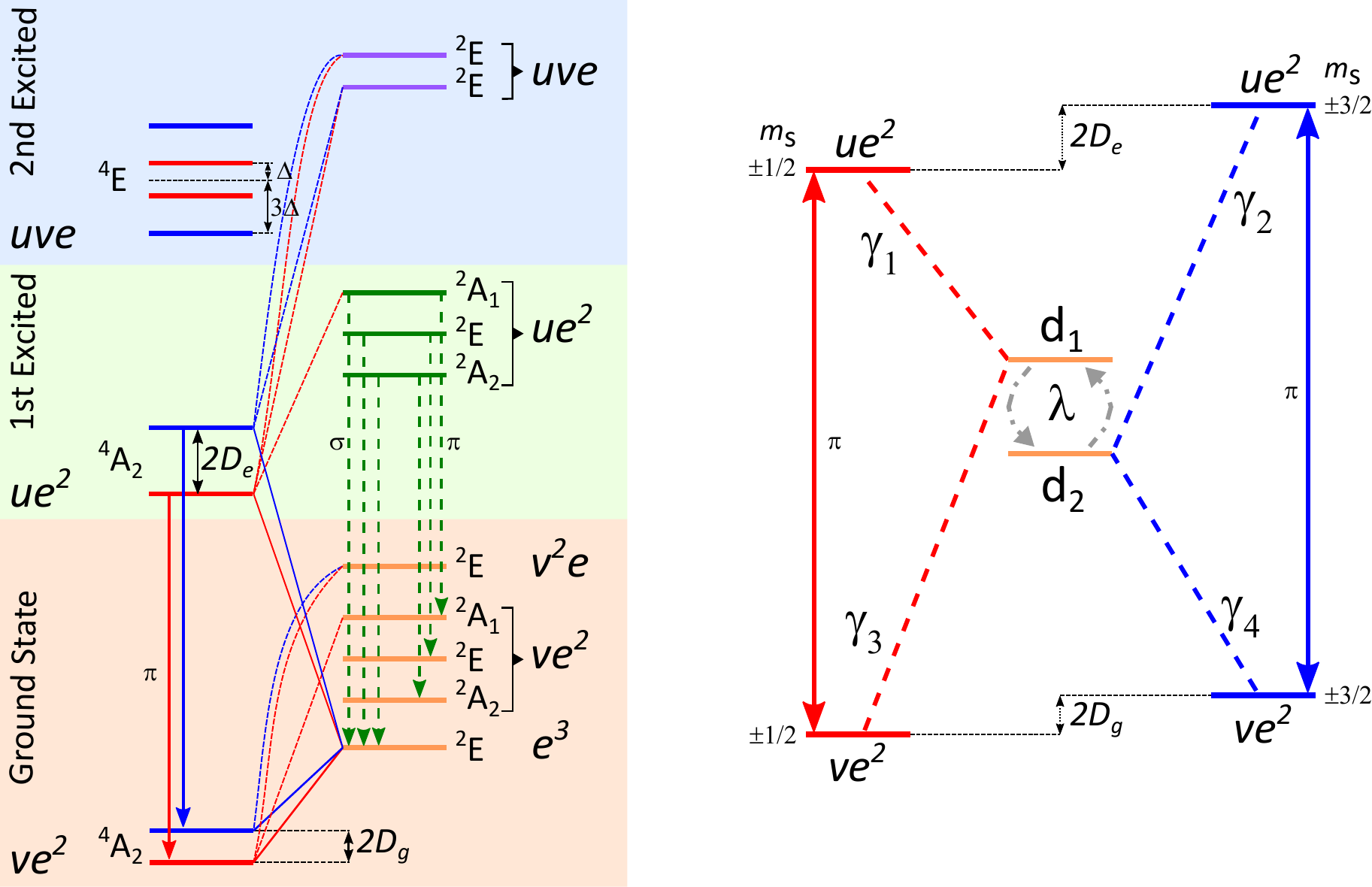}

		\caption{Full electronic fine structure of V1 (left). Simplified six-level energy scheme (right).}
		\label{fig:energylevel_theory}
\end{figure}
\end{center}

Here, we analyse the full electronic structure of the V1 defect shown in Fig.~\ref{fig:energylevel_theory} to determine the relative strengths of all spin-orbit (SO) assisted intersystem-crossing (ISC) transitions between quartet and doublet states. Ground state manifold includes $ve^2$, $v^2 e$, and $e^3$ orbital states that are energetically similar due to only slightly broken T$_\mathrm{d}$ symmetry of the vacancy center, as previously demonstrated in Ref.~\onlinecite{Soykal2016}. Inclusion of the spin degree of freedom further transforms the $ve^2$ orbital into a spin quartet (g$_1$, g$_2$) that forms the ground state of this defect and three doublet states (with A$_1$, A$_2$, and E symmetries). Each of the $v^2 e$ and $e^3$ orbital states also form E symmetry doublets. From these doublets, only $ve^2$(A$_1$), $e^3$(E), and $v^2 e$ (E) doublets are involved in the spin-selective ground-state ISC mechanism.

Similarly, excited state manifold of the defect includes $ue^2$ and $uve$ orbital states. After consideration of the spin, both $ue^2$ and $uve$ form spin quartet states that are responsible for the optical transitions; for instance, $ue^2 \rightarrow ve^2$ corresponds to the V1 transition observed at $\SI{862}{nm}$ in this work. On the other hand, $ue^2$ forms three spin doublets with A$_1$, A$_2$, and E symmetries in which only A$_1$ doublet is involved in the ISC mechanism, whereas $uve$ contributes into two E symmetry doublets both contributing to the ISC. Using the symmetry-adapted total wavefunctions of these states like those obtained in Ref.~\onlinecite{Soykal2016}, we evaluate the relative strengths of ISC transition rates (obtained through Fermi’s golden rule by using the SO coupling matrix elements) between all doublet-doublet and quartet-doublet states. We make two important observations: (i) In each ground and excited state manifold, there is a strong SO coupling among all energetically close doublets strongly hybridizing these states with each other. (ii) Radiative decay from excited to ground doublet states is most likely suppressed partly due to the mechanism outlined in (i) steering the populations from optically active doublet states toward dark doublets, favouring non-radiative decays. This is expected to manifest itself as an increased metastable lifetime of doublet states in agreement with our observations of the rates.

Following the approach and observations above, the overall spin-selective non-radiative decay rate to each ground state is obtained by adding all relative transition rates from all the doublet states of the ground-state manifold. The rates for each spin are given as $\gamma_{3} \propto (2\pi/\hbar) \{ \frac{|\lambda_{T}^2|}{3} + \frac{2|\lambda_{Z}^2|}{3} \}$ and $\gamma_{4} \propto (2\pi/\hbar) |\lambda_{T}|^2$ for $m_{s} = \pm 1/2$ and $m_{s} = \pm 3/2$, respectively, in terms of the transverse and longitudinal SO coupling parameters (with respect to the $c$-axis). We note that, due to the near-T$_\mathrm{d}$ symmetry of the defect center, SO coupling can be assumed to be almost isotropic ($\lambda_{T} \simeq \lambda_{Z}$) resulting with $\gamma_{3} \simeq \gamma_{4}$ in excellent agreement with the equal metastable state lifetime observed experimentally. Similarly, we obtain the overall non-radiative decay rate from excited state to doublet states of the excited state manifold to be  $\gamma_{1} \propto (2\pi/\hbar) \{ \frac{2|\lambda_{Z}^2|}{3} + \frac{|\lambda_{T}^2|}{9} \}$ and $\gamma_{2} \propto (2\pi/\hbar) \frac{|\lambda_{T}|^2}{3}$ for $m_{s} = \pm 1/2$ and $m_{s} = \pm 3/2$, respectively, resulting with $\gamma_{2} \simeq 0.5\gamma_{1}$ under the isotropic assumption, also in great agreement with the experimental data. From a detailed comparison of these rate expressions with those calculated later in Tab.~\ref{tab:oney_rates}, the relationship between the components of spin-orbit coupling parameters is found to be $\lambda_{Z} \leq  \lambda_{T} \leq 1.13 \lambda_{Z}$ within the error margin of our measurements and simulations. This further validates the full electronic structure model outlined here involving the V1 center’s ISC mechanism.

The fluorescence and ISC mechanism can be greatly simplified into a six-level model of Fig.~1(b) of the main text and Fig.~\ref{fig:energylevel_theory}, which was obtained in Ref.~\onlinecite{Nagy2019}, while also keeping the strong spin mixing among the hybridized doublet states as an adjustable parameter $\lambda$. Note that, in this simplified model, the decay times from excited state doublets to ground state doublets present in the full model is contained in the metastable lifetime of the doublet states. For both $A_1$ and $A_2$ transitions corresponding to spin-1/2 and spin-3/2 subspaces, the radiative decay rates including zero-phonon-line (ZPL) and phonon sidebands (PSBs) are the same. The latter is directly understood from the fact that PSB radiative transitions are the interplay between the orbital wavefunctions of the defect and phonons of the host materials. Both spin 1/2 and 3/2 have the same orbital wavefunctions belonging to the same symmetry group. The phonon modes of 4H-SiC are also spin-independent. Thus, the phonon assisted processes are naturally spin-independent.

\section{Estimation of the non-radiative decay rates associated with the $A_1$ and $A_2$ transitions\label{sec:phonon}}
The non-radiative decay rates $\Gamma_{1,2}$ of the V1 defect are governed by a high-order electron-phonon coupling that involves multi-phonon relaxation of the excited state to ground state. This high-order electron phonon coupling term dominates over the linear electron-phonon coupling term, treated as a high order perturbation, whenever the energy gap between the electronic states ($\omega_0$) greatly exceeds the highest optical phonon mode frequency ($\omega_\mathrm{op}$) available in the host material. This happens to be the case for the $A_1$ and $A_2$ transitions of the V1 center with an energy gap of $\omega_0 = \SI{1.438}{\textit{e}V}$  between excited and ground states, and highest phonon cut-off frequency of $\omega_\mathrm{op} = \SI{118.3}{m\textit{e}V}$  at $\Gamma$-point for bulk 4H-SiC revealing that this direct relaxation mechanism is governed by 12$^\mathrm{th}$ order electron-phonon coupling involving minimum of 12 optical phonons. Therefore, it is expected to have a much slower decay rate than its radiative counterpart. In addition, even though the multi-phonon relaxation rate decreases exponentially with an increased energy gap, the energy difference between $A_1$ and $A_2$ optical transitions ($\sim \SI{4}{\upmu  \textit{e}V}$) is significantly smaller than $\hbar \omega_\mathrm{op}$ ($\sim\SI{118}{m\textit{e}V}$). Therefore, it is reasonable to assume $\Gamma_{1} = \Gamma_{2} = \Gamma$. Since the main purpose here is only to provide an order of magnitude estimate for the upper bound of $\Gamma_{1,2}$, instead of using the localized phonon mode frequency in the presence of the V1 center, one can use the bulk cutoff frequency. This is further rationalized as the energies of these localized (flat-band) phonon modes, introduced by the presence of the V1 center, are expected to be very close (only slightly higher) to the bulk optical phonon cutoff energy, as also recently observed for the V2 centers in Ref.~\onlinecite{Shang2020}.

The non-radiative decay rate involving multi-phonons is given (see Ref.~\onlinecite{Burstein2010}) by $\Gamma_{i} = \kappa e^{-\alpha \hbar  \omega_0}$ in which $\alpha = \ln (D/\hbar \omega_\mathrm{op}) / (\hbar \omega_\mathrm{op})$  and $\kappa = f B(p) [(N_{c} -1)/N_{c}^{2}] (4 \pi ^2 \rho_\mathrm{M} a^3 D )/(3\hbar m)$ mainly depending on the atomic configuration of the defect and the properties of the host material. The parameters $D, \rho_\mathrm{M},m,a$ correspond to the bonding dissociation energy of the defect, material density ($\SI{3.17}{gcm^{-3} }$), electron mass, and average lattice constant ($\sim \SI{3.094}{\AA}$), respectively. $N_{c} = 4$ is for the number of atoms contributing to the atomic basis of the vacancy center. The function $B(p)$ originates from the Taylor expansion term belonging to the $p^\mathrm{th}$-order vibrational mode within an equivalent Morse potential. It is defined analytically as $B(p)=(2^{p}-1)^2 p^{(p-2)}/([\Gamma (p+1)]^{2})$ in terms of the Gamma-function and depends only on the minimum number of phonons involved ($p=\omega_0/\omega_\mathrm{op} \simeq 12$) in the relaxation through the energy gap resonantly. For a vacancy center, the dissociation energy is defined as $D=(E_{f}+E_{d}+E_{r} )/2$ (see Ref.~\onlinecite{Larkins1971}) using a molecular orbital approach in whi{c}h $E_{f}$ is the formation energy of the negatively charged vacancy. Second contribution $E_{d}$ is the difference in energy between the configuration in which the five active electrons distributed in each of the four hybridized ($sp^3$) orbitals with random spin orientation and the configuration of electrons in the ground state molecular orbital ($ve^2$) for the undistorted system. A third small correction $E_{r}$ can be added for the lowering of energy due to the relaxation of atoms near the vacancy center. Since we are only looking to estimate an upper bound for the non-radiative decay rate here, we use a conservative smallest possible value for $D$ given as the half of the formation energy $E_{f}$ ($D = E_{f}/2 = \SI{4.48}{\textit{e}V}$) calculated for the negatively charged V1 center (see Ref.~\cite{Bockstedte2003}). Overall, the non-radiative lifetime depends exponentially on the number ($p$) of cutoff phonons involved in the relaxation process, but slightly scale with the inverse of the oscillator strength $f^{-1}$. The oscillator strength can be estimated from $f=(6\pi \epsilon_0 mc^3 )/(n^3 e^2 \omega_0 ^2 \tau_\mathrm{r})$ using the experimentally obtained total radiative lifetime, $\tau_\mathrm{r} \leq \SI{21}{ns}$. From this, we estimate the non-radiative decay rate to be $\Gamma_{1,2} \leq (\SI{21}{ms})^{-1}$, six orders of magnitude slower than its radiative counterpart. Last, the temperature dependence of the non-radiative decay rate is defined as $\Gamma_{1,2}(T) = \Gamma_{1,2}[  1+ 1/(e^{\hbar\omega_\mathrm{op}/k_\mathrm{B}T} -1 ) ]^{p}$ showing almost no temperature dependence up to room temperature.

\section{Pulse-train measurement: theory of initial ground-state spin populations prepared by a train of short pulses\label{sec:theory_pulsetrain}}
The key to precise determination of the rates is the preparation of the initial spin population well defined by the rates of the spin system. Optical pumping and the intersystem-crossing mechanism are widely used for the spin-state preparation. The optical pumping can be performed with either continuous-wave (cw) laser or with pulsed laser, but spin-state prepared by cw laser can be laser power dependent. As the excitation strength can fluctuate due to the laser and focus instability etc., reliable and reproducible spin-state preparation is not very easy with cw laser. Instead in this study, we use a train of short pulses whose pulse interval $t_\mathrm{p}$ (denoted as $t_\mathrm{init}$ in the main text) is much longer than the metastable-state lifetime to initialize the spin population. This method allows for excitation-strength-independent preparation of the spin state, and thus the spin polarization is ideally determined solely by the rates of the system.

First, we derive the analytic formula of the ground-state spin population by the pulse-train method based on the five-level rate model shown in Fig. 1(a) in the main text. Let the ground, excited, and metastable-state spin population at time $t$ after $k$-th pulsed excitation be $n_{\mathrm{g}_{i}}^{k}(t)$, $n_{\mathrm{e}_{i}}^{k}(t)$ and $n_\mathrm{ms}^{k}(t)$, respectively, where $i=1,2$ denotes the spin sublevels $\ket{\pm 1/2}$ and $\ket{\pm 3/2}$, respectively. By solving the rate equations, the spin populations after $k$-th pulsed excitation are obtained as
\begin{gather*}
n_{\mathrm{g}_{i}}^{k} (t)  = (1-P_\mathrm{e}) n_{\mathrm{g}_{i}}^{k-1} (t_\mathrm{p}) + (\tau_{\mathrm{e}_{i}}^{-1}  - \gamma_{i}) \int_{0}^{t} n_{\mathrm{e}_{i}}^{k}(t^{\prime}) \mathrm{d} t^{\prime} + \frac{\gamma_{i+2}}{2}  \int_{0}^{t} n_{\mathrm{ms}}^{k}(t^{\prime}) \mathrm{d} t^{\prime},\\
n_{\mathrm{e}_{i}}^{k}(t)  = [n_{\mathrm{e}_{i}}^{k}(t_\mathrm{p}) + P_\mathrm{e} n_{\mathrm{g}_{i}}^{k}(t_\mathrm{p})  ]e^{-t/\tau_{\mathrm{e}_i} },	\\
n_{\mathrm{ms}}^{k} (t)  = n_{\mathrm{ms}}^{k-1} (t_\mathrm{p}) e^{ -\gamma_\mathrm{ms}t  } + \sum\limits_{i=1}^{2} \frac{  \gamma_{i}\tau_{\mathrm{e}_{i}}  n_{\mathrm{e}_{i}}^{k}(0)      }{1- \gamma_\mathrm{ms} \tau_{\mathrm{e}_{i}}} [e^{ -\gamma_\mathrm{ms}t  } - e^{ -t/\tau_{\mathrm{e}_{i}}  }] ,
\end{gather*}
where $\gamma_\mathrm{ms} = (\gamma_{3}+\gamma_{4})/2$. After a large number of pulses ($k\rightarrow \infty$) with a long pulse interval $t_\mathrm{p} \gg \tau_\mathrm{ms} \gg \tau_{\mathrm{e}_{i}}$, the ground-state populations reach a steady state and are calculated to be
\begin{align}
	\begin{split}
	\label{eq:init_ground}
	n_{\mathrm{g}_1}^{\infty} (\infty) & = \frac{\gamma_{2}\tau_{\mathrm{e}_{2}} / \gamma_{4}}{ \gamma_{1}\tau_{\mathrm{e}_{1}}/\gamma_{3} + \gamma_{2}\tau_{\mathrm{e}_{2}}/\gamma_{4}},\\
	n_{\mathrm{g}_2}^{\infty} (\infty) & = \frac{\gamma_{1}\tau_{\mathrm{e}_{1}} / \gamma_{3}}{ \gamma_{1}\tau_{\mathrm{e}_{1}}/\gamma_{3} + \gamma_{2}\tau_{\mathrm{e}_{2}}/\gamma_{4}}.
	\end{split}
\end{align}
These results suggest that the spin polarization with a train of short pulses is independent of the excitation strength and well defined only by the excited-state lifetime and the intersystem crossing rates. The Eq.~\ref{eq:init_ground} is valid even when the direct non-radiative decay channel ($\Gamma_{1,2}=\Gamma$ in the main text) exists.

\begin{center}
	\begin{figure}[h!]
		\includegraphics[trim=0 0 0 0,clip,scale=0.55]{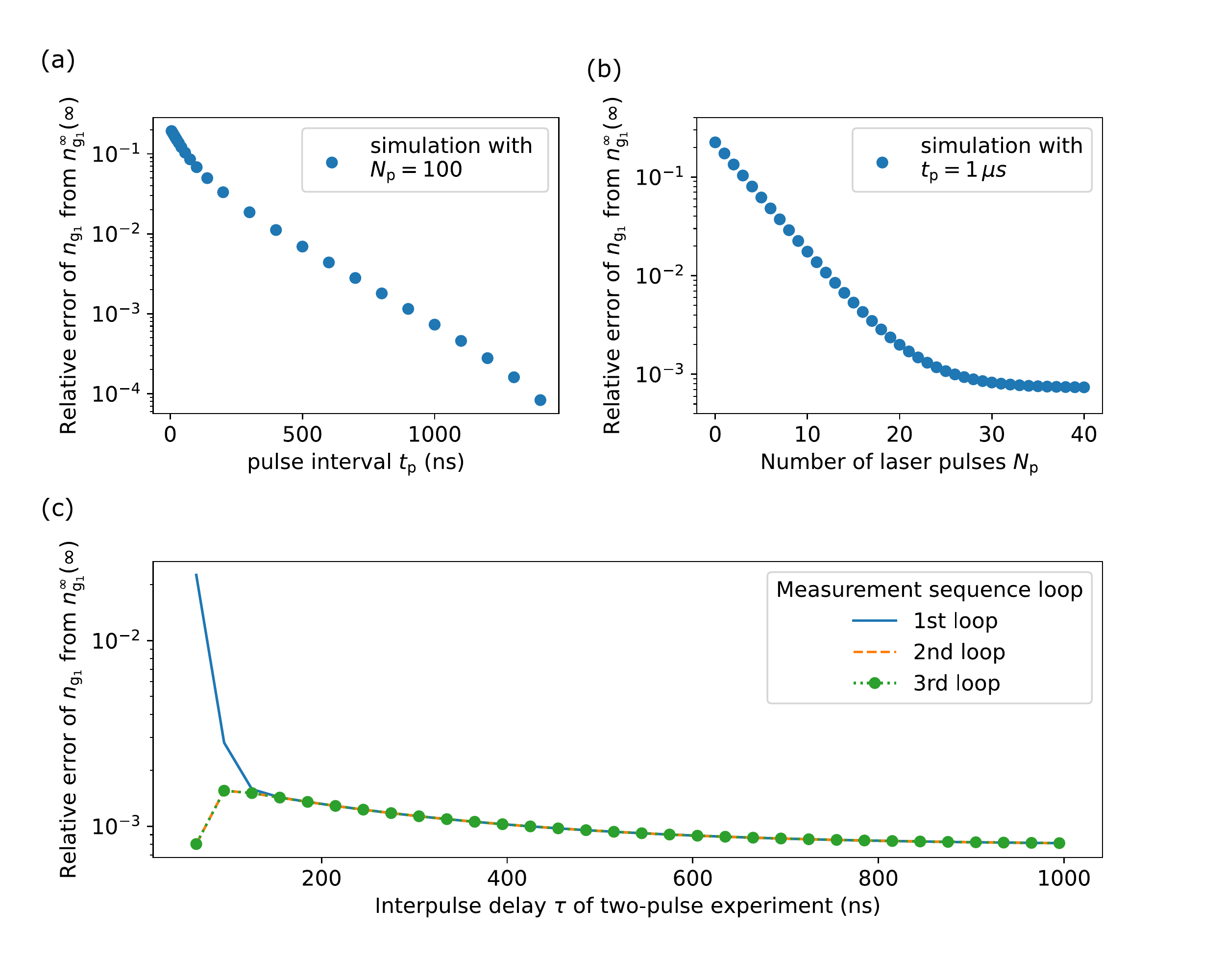}
		\caption{Simulation of the initialization of the ground-state spin population using a pulse train of short pulses. (a) Spin population prepared with various pulse interval $t_\mathrm{p}$ with large number of initialization laser pulses $N_\mathrm{p}=100$. (b) Ground-state spin population in $m_{s}\pm 1/2$ prepared with various number of initialization laser pulses at fixed $t_\mathrm{p} = \SI{1}{\upmu s}$. The initial state (before applying any laser pulse) is in the completely depolarized state. (c) Simulated spin population in $m_{s}\pm 1/2$ with 9 initialization pulses at each inter-pulse delay in the pulse-train experiment at first, second, and the third round of the measurement sequence. The spin polarization can be well prepared from the second round of the sequence. For all plots, $P_\mathrm{e}=0.608$ was used, and the vertical axis is a relative error of the spin population in $m_{s}\pm 1/2$ from the ideal theoretical value of $n_{\mathrm{g}_1}^{\infty} (\infty)$ given in Eq.~\ref{eq:init_ground}.}
		\label{fig:naoya_theory}
	\end{figure}
\end{center}

In experiments, the pulse interval $t_\mathrm{p}$ and the number of the initialization pulses $N_\mathrm{p}$ cannot be infinite. A longer $t_\mathrm{p}$ and a larger $N_\mathrm{p}$ are preferable, but these values have to be determined to keep the measurement time realistic. In our experiment, we used $t_\mathrm{p}= \SI{1}{\upmu s}$ and $N_\mathrm{p}=9$. After nine initialization pulses, we waited for $\SI{2}{\upmu s}$ to ensure that the system decays back to the ground state. The ground-state spin population at this point is the initial state for the two-pulse measurement. The pulse interval of $t_\mathrm{p}= \SI{1}{\upmu s}$ ($\simeq 4 \tau_\mathrm{ms}$) was determined by simulating the $t_\mathrm{p}$ dependence of the ground spin population, which is shown in Fig.~\ref{fig:naoya_theory}(a). With this condition, the spin population is prepared with a precision with an error smaller than 0.1\%. At $t_\mathrm{p}= \SI{1}{\upmu s}$, we also simulated the $N_\mathrm{p}$ dependence of the ground state population and the result is displayed in Fig.~\ref{fig:naoya_theory}(b). More than 20 pulses are necessary to prepare the spin state into $n_{\mathrm{g}_1}^{\infty} (\infty)$ with a precision of 0.1\% error. Nevertheless, we show that the pulse-train experiment can be performed with smaller $N_\mathrm{p}$ with a good precision. 

In the pulse-train measurement, two laser pulses with a pulse interval $\tau_{m}$ are applied after spin state preparation followed by a 1-$\mathrm{\upmu s}$-long decay time before the next initialization for the measurement at $\tau_{m+1}$. The pulse sequence $S_{m}$ for the measurement at $\tau_{m}$ thus consists of \{9 initialization pulses with $\SI{1}{\upmu s}$ interval, $\SI{2}{\upmu s}$ decay time, pump laser pulse, $\tau_{m}$ delay, probe laser pulse, $\SI{1}{\upmu s}$ decay time\}, and we sequentially measured 32 points of $\tau_{m}=(65+30m)\,\mathrm{ns}$, i.e. for $m=0,1,\cdots,31$. After $S_{31}$, we waited for $\SI{100}{ns}$ to repeat the sequence from $S_0$ again. Therefore, one loop of the whole sequence $S_\mathrm{tot}$ is $\{S_0, S_1, \cdots, S_{31}, \SI{100}{ns} \ \mathrm{wait} \}$. Although the interval of the laser pulses is perturbated every 10 pulses due to varying $\tau_{m}$, the sequence $S_\mathrm{tot}$ basically consists of pulse trains with very large number of laser pulses. Therefore, the spin population is expected to be well initialized during the first loop of the sequence. To demonstrate this, we simulated the whole sequence $S_\mathrm{tot}$ at the first, second, and the third loop of the measurement, and studied how the ground-state spin population changes over the repetition of the sequence. Fig.~\ref{fig:naoya_theory}(c) shows the simulated ground-state spin population in terms of the difference from the theoretical population $n_{\mathrm{g}_1}^{\infty} (\infty)$ in $m_{s}\pm 1/2$ right before the 'pump' pulses at each $\tau_{m}$. Although the spin population largely deviates from $n_{\mathrm{g}_1}^{\infty} (\infty)$ in the beginning of the first loop, the deviation becomes less than 0.2\% at all $\tau_{m}$ from the second loop of the sequence. The pulse interval perturbation by $\tau_{m}$ introduces a small deviation of the spin population from the steady state, but it is well suppressed by 9 initialization pulses. In the experiment, we record the data 1 second after initiating the sequence.

\section{Pulsed excitation probability and pulse energy fluctuation\label{sec:appendix4}}
\begin{center}
	\begin{figure}[h!]
		\includegraphics[trim=0 0 0 0,clip,scale=0.7]{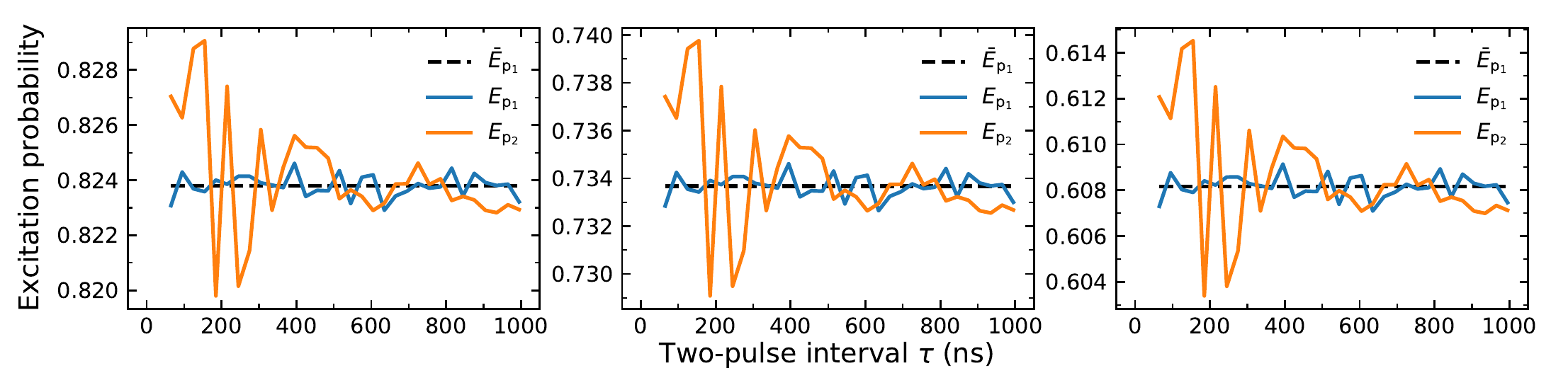}
		\caption{Pulse energy fluctuation of the 'pump' and 'probe' pulses in the pulse-train measurement is shown for different excitation probabilities in panels (a), (b) and (c).}
		\label{fig:pulse_fluc}
	\end{figure}
\end{center}
The fluorescence count rate $(I)$ from a single defect after the background subtraction follows an exponential function of the laser power \cite{CHAPMAN2011} given by
\begin{equation}\label{eq:pulsed_saturation}
	I = I_0 \times \left(1-e^{     -E_\mathrm{p}/E_\mathrm{s}   } \right),
\end{equation}
where $I_0$ and $E_\mathrm{s}$ are the pulse energy and count rate at saturation respectively. The pulse energy $E_\mathrm{p}$ is obtained from the measured power $P$ and the pulse repetition rate $R$ by $E_\mathrm{p} = P/R$. From Eq.~\ref{eq:pulsed_saturation} the excitation probability of each laser pulse is derived to be
\begin{equation}\label{eq:exc_prob}
	P_\mathrm{e} = 1-e^{-E_\mathrm{p}/E_\mathrm{s} }
\end{equation} 
To check the energy fluctuation of the pulsed laser employed in the pulse-train excitation scheme, we measured the laser photon counts ($\sim$ pulse energy $E_\mathrm{p}$) as a function of the pulse interval ($\tau$) between the 'pump' and 'probe' pulses. Equation (3) in the main text employs the averaged pulse energy with regard to the first excitation pulse ('pump'), i.e. $P_\mathrm{e,avg}=1-e^{-\bar{E}_\mathrm{p_{1}}/E_\mathrm{s} }$ (dashed lines in Fig.~\ref{fig:pulse_fluc}). The actual excitation probabilities considering the pulse energy fluctuation are also illustrated in Fig.~\ref{fig:pulse_fluc}) for the 'pump' and 'probe' pulses. The overall fluctuation of the excitation probability is only within 1\%. The excitation probability was corrected using the measured pulse energy fluctuation in the analysis of the pulse-train measurement. 

\section{Resonant spin-polarization readout scheme proposed for $\boldsymbol{\gamma_{3} \neq \gamma_{4}}$\label{sec:appendix5}}
\begin{center}
	\begin{figure}[h!]
		\includegraphics[trim=0 0 0 0,clip,scale=0.75]{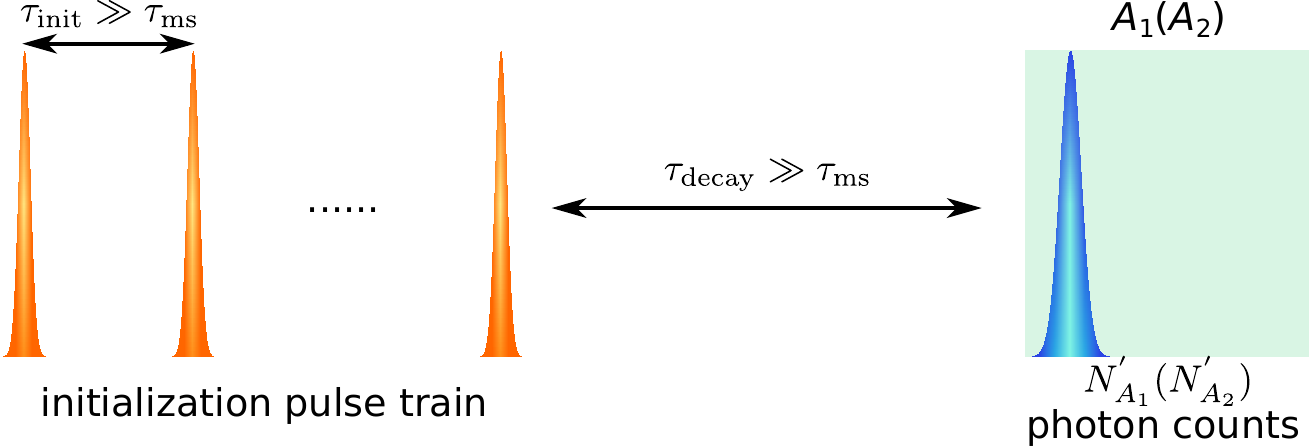}
		\caption{Measurement sequence for resonant spin-polarization readout.}
		\label{fig:res_spin_readout}
	\end{figure}
\end{center}
The intersystem-crossing (ISC) rates out of the metastable state to the ground states ($\gamma_{3}$ and $\gamma_{4}$) can be differentiated by a resonant readout scheme as shown in Fig.~\ref{fig:res_spin_readout}. The ground-state spin populations of $n_{\mathrm{g}_1}$ and $n_{\mathrm{g}_2}$ are initialized into the same states in Eq.~\ref{eq:init_ground} by the same above-resonant pulse train (see Appendix \ref{sec:theory_pulsetrain}) and subsequently read out by the $A_1$ and $A_2$ pulsed resonant lasers separately. The ratio of fluorescence integrated over a full-time window probed by $A_1$ and $A_2$ resonant lasers depends on all the ISC rates
\begin{equation}\label{eq:res_spin_readout}
	\frac{N_{\mathrm{A_1}}^{\prime}}{N_{\mathrm{A_2}}^{\prime}} = \frac{\gamma_{2}/\gamma_{1}}{\gamma_{4}/\gamma_{3}}
\end{equation}
Together with the experiments in the main text, $\gamma_{3}$ and $\gamma_{4}$ can be distinguished using this resonant readout scheme.

\section{Resonant depletion measurement: quasi-continuous wave initialization\label{sec:appendix6}}
The resonant depletion measurement scheme requires cw resonant initialization and subsequent readout by resonant pulses. To realize this within one resonant excitation path (see Fig.~2 in the main text), we initialized instead using quasi-cw excitation in $\SI{20}{MHz}$-sine intensity waveform. The intensity output of an electro-optic amplitude modulator (EOAM) is given by
\begin{equation}\label{eq:intensity_mod}
	I = I_0 \times I_\mathrm{mod}=I_0 \sin ^{2} \left(  \frac{\pi V_\mathrm{mod}}{V_{\pi}}   \right)
\end{equation}
where $V_{\pi}$ is the $\pi$-voltage of the EOAM, $V_\mathrm{mod}$ the modulation voltage. The unmodulated light intensity which contains the transition frequency at resonance is expressed by $I_0 \sim \sin^{2}(2\pi f_0 t) \sim E^{2} $. Since the amplitude of the $\SI{10}{MHz}$ sine modulation voltage ($V_\mathrm{a}$) is much smaller than the $\pi$-voltage $V_\mathrm{a} \ll V_\mathrm{\pi}$, we have $\sin^{2} \left(  \frac{\pi V_\mathrm{mod} }{V_{\pi}}  \right) \approx \left( \frac{\pi V_\mathrm{a} \sin (2\pi ft) }{V_{\pi}}  \right)^{2} \sim \sin^{2} (2\pi ft)$.Thus, the modulated electric field of the resonant laser is $E_\mathrm{mod} =\sqrt{I_\mathrm{mod}}\sim |\sin (2\pi ft)|$ with $f=\SI{10}{MHz}$. With trigonometric identities, we see from Eq.~\ref{eq:intensity_mod} that the modulated electric field contains sidebands at $f_0 \pm \SI{10}{MHz}$. In other words, the bandwidth of the resonant laser is $\SI{20}{MHz}$ which is within the absorption linewidth of V1 measured in Refs.~\cite{Nagy2019,Morioka2020}. Thus, the cw resonant initialization is well approximated by the quasi-cw excitation under the $\SI{20}{MHz}$-sine intensity modulation.

\section{Parameter fine-tuning: extracting the rates from the resonant depletion measurement\label{sec:appendix7}} 
The Hamiltonian of the simplified six-level electronic structure of V1 (see Appendix \ref{sec:electronic_structure}) is given by
\begin{gather*}
	H = \frac{2D_\mathrm{g} - 2D_\mathrm{e} + \delta_\mathrm{L}}{2} ( \ket{\mathrm{g_1}}\bra{\mathrm{g_1}} - \ket{\mathrm{e_1}}\bra{\mathrm{e_1}}  ) - \frac{2D_\mathrm{g} - 2D_\mathrm{e} - \delta_\mathrm{L}}{2} ( \ket{\mathrm{g_2}}\bra{\mathrm{g_2}} - \ket{\mathrm{e_2}}\bra{\mathrm{e_2}}  ) \\
	+ \lambda ( \ket{\mathrm{d_1}}\bra{\mathrm{d_2}} + \ket{\mathrm{d_2}}\bra{\mathrm{d_1}} ) + \Omega ( \ket{\mathrm{g_1}}\bra{\mathrm{e_1}} +  \ket{\mathrm{g_2}}\bra{\mathrm{e_2}} +h.c.)
\end{gather*}
\noindent
in the rotating frame of the resonant laser in terms of ground ($\mathrm{g_1}$, $\mathrm{g_2}$), excited ($\mathrm{e_1}$, $\mathrm{e_2}$), and doublet $\mathrm{d_1}$, $\mathrm{d_2}$ states for each spin multiplicity. The $D_\mathrm{g}$ and $D_\mathrm{e}$ are the ground and excited state zero-field splitting parameters, respectively. The laser detuning is introduced by $\delta_\mathrm{L}$ that allows addressing each $A_1$ and $A_2$ optical transition resonantly with a Rabi frequency of $\Omega$. The Lindblad master equation is defined as 
\begin{equation*}
	\frac{\partial \rho}{\partial t} = -i [H, \rho] + \gamma_\mathrm{r} [L(\Lambda_\mathrm{r_1})+ L(\Lambda_\mathrm{r_2})   ] + \sum\limits_{i=1}^{4} \gamma_{i}L(\Lambda_{i}) + \gamma_{s}L(\Lambda_{s})
\end{equation*}
using $L(\Lambda_{i}) = \Lambda_{i}\rho \Lambda_{i}^{\dagger} - \{\Lambda_{i}^{\dagger}\Lambda_{i}, \rho    \}/2$ in which $\Lambda_{i}$ are the transition operators that correspond to each $\gamma_{i}$ decay channel, e.g. $\Lambda_{1} = \ket{\mathrm{d_1}}\bra{\mathrm{e_1}} $ for $\gamma_{1}$. Moreover, additional pure spin dephasing is included with $\Lambda_{s} = \ket{\mathrm{d_1}}\bra{\mathrm{d_1}} - \ket{\mathrm{d_2}}\bra{\mathrm{d_2}}$ operator.

\begin{center}
	\begin{figure}[h!]
		\includegraphics[trim=0 0 0 0,clip,scale=0.9]{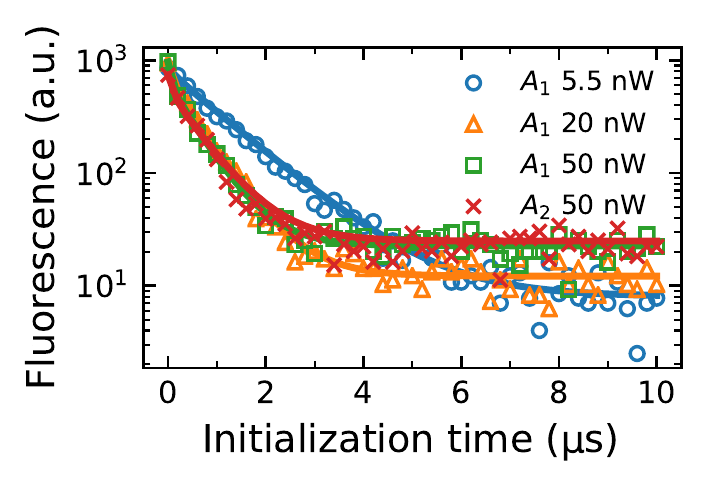}
		\caption{Paramater fine-tuning based simulated resonant depletion curves denoted by solid lines.}
		\label{fig:dep_Oney}
	\end{figure}
\end{center}

The resulting time-dependent density matrix master equation is parametrized in terms of unknown rates and solved to simulate the fluorescence from $A_1$ and $A_2$ transitions. The rates given in Tab.~\ref{tab:oney_rates} are obtained from the fitting of these results to the resonant depletion measurements at different initialization powers simultaneously using a custom-built Nelder-Mead and differential evolution optimization algorithm over a large space of $\gamma_{i}$, $\gamma_{s}$, $\gamma_\mathrm{r}$ values. To ensure the algorithm is not stuck at a local minimum, we performed convergence analysis over $\gamma_{1}$ and $\gamma_{2}$ after each iteration. Conservative estimates of the error margins of these rates are also provided in Tab.~\ref{tab:oney_rates}, and they are mostly due to the slight scattering present in the experimental fluorescence data points, with varying degrees for different powers, limiting the fitness of the solutions within a given confidence interval (95\%). The resulting simulated fluorescence of the $A_1$ and $A_2$ optical transitions using the rates we obtained are in excellent agreement with the experimental values as shown in Fig.~\ref{fig:dep_Oney}.
\begin{table}
	\centering
	\caption{Summary of radiative rates, ISC rates and MS lifetime for single V1 centers obtained by the fitting to the measured resonant depletion data.}
	\label{tab:oney_rates}
	\begin{tabular}{cc}
		\hline\hline
		& Parameter fine-tuned (ns)\\\hline
		e$_{1,2}\rightarrow$ g$_{1,2}$: $\gamma_\mathrm{r}^{-1}$ & $9.1\pm 0.2$  \\ 
		e$_{1} \rightarrow$ d: $\gamma_{1}^{-1}$ & $11.3\pm 0.3$  \\
		e$_{2} \rightarrow$ d: $\gamma_{2}^{-1}$ & $20.6\pm 1.1$  \\
		d $\rightarrow$ g$_1$: $\gamma_{3}^{-1}$ & $270\pm 10$ \\
		d $\rightarrow$ g$_2$: $\gamma_{4}^{-1}$ & $250\pm 10$ \\
		\hline\hline  
	\end{tabular}
\end{table} 

\section{Estimation of the local electric field at the defect in the SIL\label{sec:appendix8}}
\begin{center}
	\begin{figure}[h!]
		\includegraphics[trim=0 0 0 0,clip,scale=0.5]{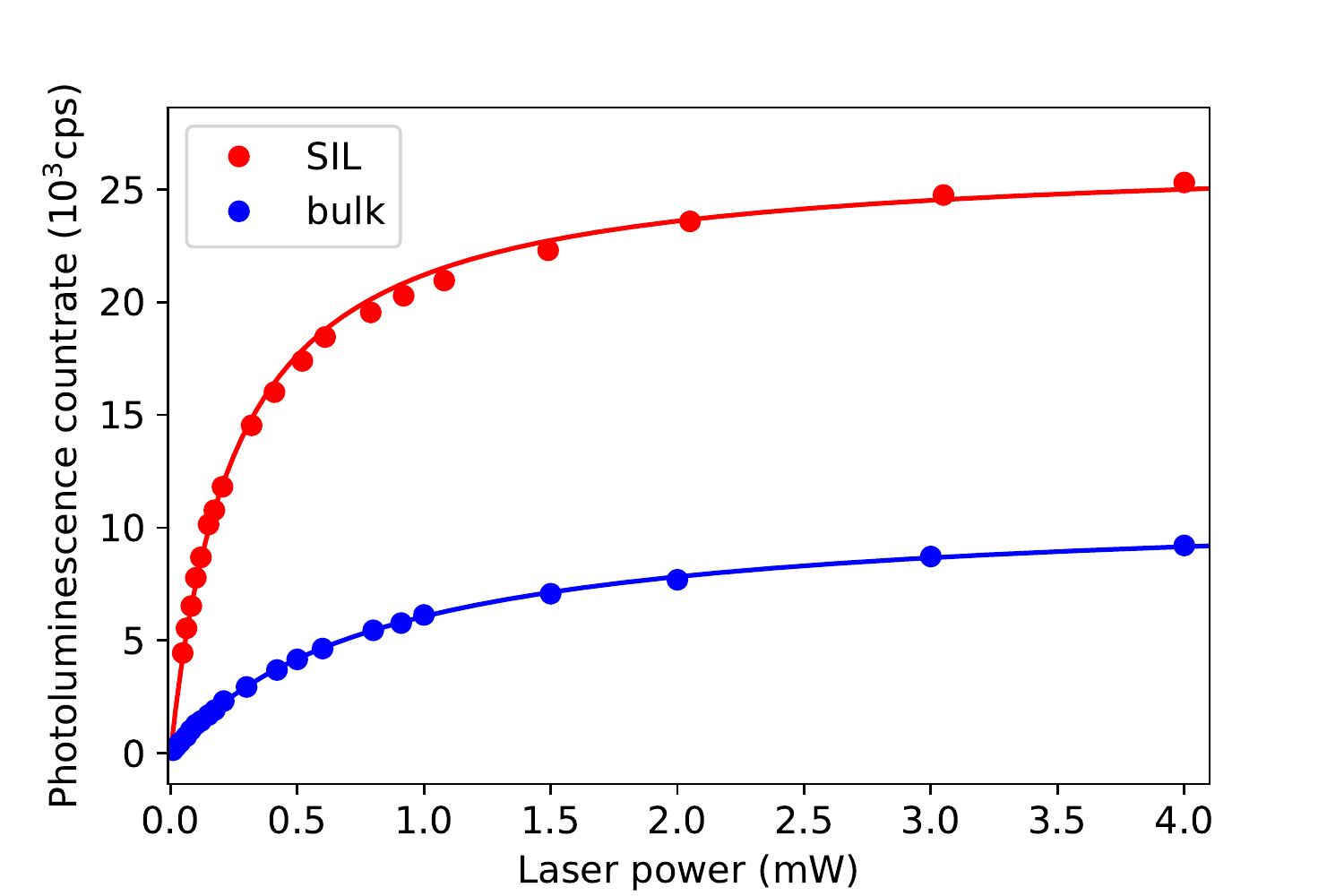}
		\caption{The photoluminescence as a function of the laser power (cw $\SI{730}{nm}$) measured before the objective for the defect in SIL (on which all the measurements mentioned in the main text were conducted) and a V1 defect in the bulk area.}
		\label{fig:saturation}
	\end{figure}
\end{center}

\begin{center}
	\begin{figure}[h!]
		\includegraphics[trim=0 0 0 0,clip,scale=0.5]{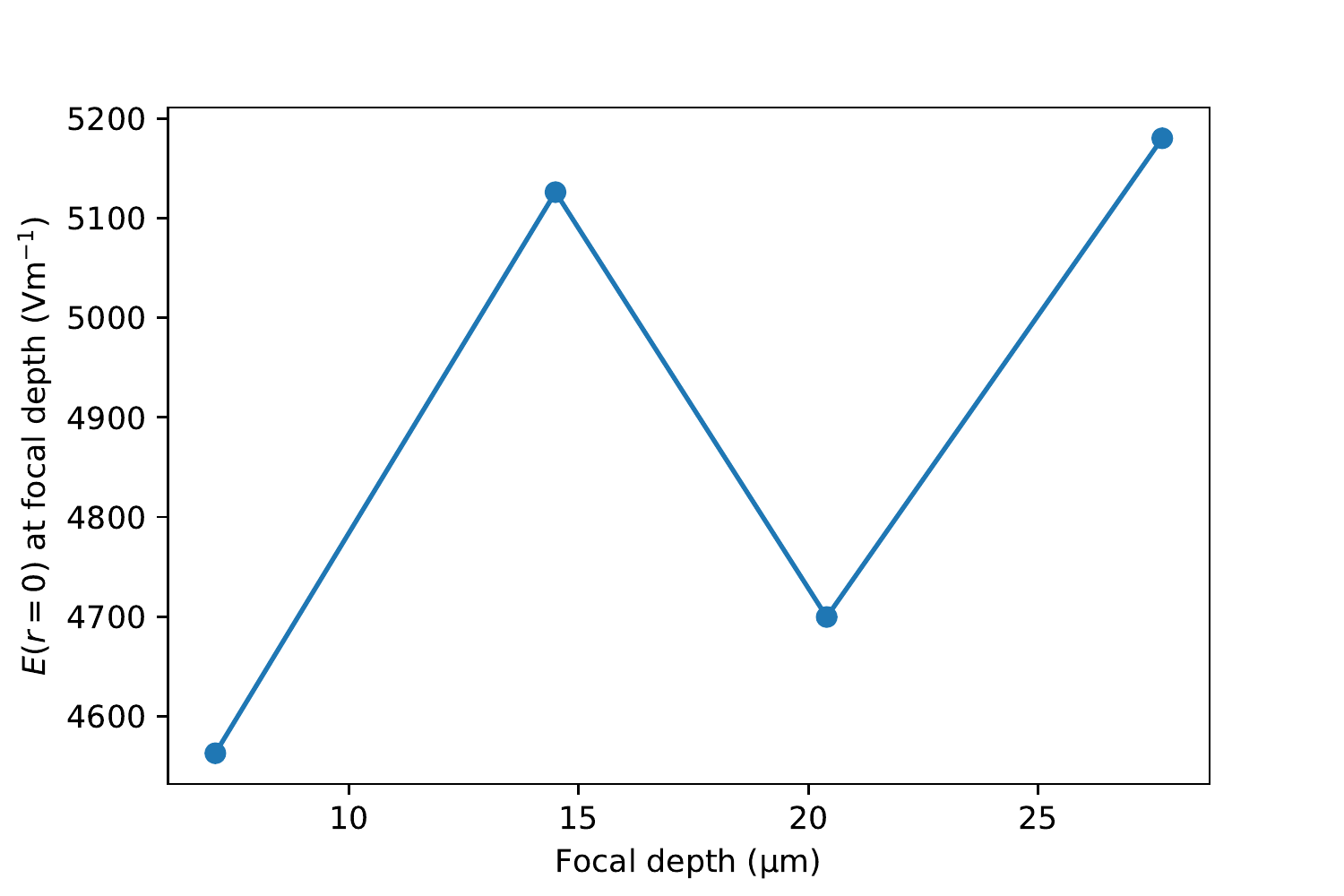}
		\caption{The electric-field strength at focus as a function of the focal depth, i.e. the distance from the focus to the air-SiC(bulk) interface simulated from FDFD method using the diffraction limited beam waist ($\mathrm{NA}=0.9$) and the measured $\pi$-pulse power as input parameters. }
		\label{fig:focal_depth}
	\end{figure}
\end{center}

The $\pi$-pulse energy of the temporal Gaussian pulse with an intensity FWHM $\SI{1.5}{ns}$ is $\SI{2.8}{fJ}$ (see Fig.~3(d) in the main text). Considering the transmission of the objective (Zeiss) at $\SI{862}{nm}$ is $\sim87\%$, the peak power of the temporal Gaussian profile corresponding to the $\pi$-pulse is calculated to $P_\mathrm{0,max}^{\pi} = \SI{1.75}{\upmu W} \times 87\% = \SI{1.52}{\upmu W}$. Fig.~\ref{fig:saturation} depicts the saturation measurement of V1 in SIL and bulk regions, where the electric-field enhancement in SIL is related to the ratio of the saturation powers (see the main text). The saturation powers are $P_\mathrm{sat,SIL}=\SI{254}{\upmu W}$ and $P_\mathrm{sat,bulk}=\SI{819}{\upmu W}$ from fits. The electric-field strength in the SiC bulk is inferred from the finite-difference frequency-domain (FDFD) simulation by approximating the incident focusing laser as a spatial Gaussian beam. The smallest beam waist by an objective with $\mathrm{NA}=0.9$ is given as $\omega_0=\SI{245}{nm}$ for a wavelength of $\SI{862}{nm}$ at the diffraction limit. In the simulation, $\omega_0$ and $P_\mathrm{0,max}^{\pi}$ are the input parameters of the Gaussian beam. The depth (beneath the air-SIL interface) of the defect in SIL is in the order of $\sim\SI{10}{\upmu m}$ and the measured bulk defect is located $\SI{10}{\upmu m}$ to $\SI{20}{\upmu m}$ beneath the air-bulk interface. At these depths, the relative variation of the electric-field strength in the bulk from the simulation is within 5\% as depicted in Fig.~\ref{fig:focal_depth}. Thus, the electric field at focus in bulk is $E_\mathrm{bulk} = \SI{4.9}{kV\cdot m^{-1}}$. Note that the transition dipole moment estimated from this field strength is a lower limit as it is inferred from the diffraction limited beam waist.

\end{appendix}


\begin{thebibliography}{47}%
	\makeatletter
	\providecommand \@ifxundefined [1]{%
		\@ifx{#1\undefined}
	}%
	\providecommand \@ifnum [1]{%
		\ifnum #1\expandafter \@firstoftwo
		\else \expandafter \@secondoftwo
		\fi
	}%
	\providecommand \@ifx [1]{%
		\ifx #1\expandafter \@firstoftwo
		\else \expandafter \@secondoftwo
		\fi
	}%
	\providecommand \natexlab [1]{#1}%
	\providecommand \enquote  [1]{``#1''}%
	\providecommand \bibnamefont  [1]{#1}%
	\providecommand \bibfnamefont [1]{#1}%
	\providecommand \citenamefont [1]{#1}%
	\providecommand \href@noop [0]{\@secondoftwo}%
	\providecommand \href [0]{\begingroup \@sanitize@url \@href}%
	\providecommand \@href[1]{\@@startlink{#1}\@@href}%
	\providecommand \@@href[1]{\endgroup#1\@@endlink}%
	\providecommand \@sanitize@url [0]{\catcode `\\12\catcode `\$12\catcode
		`\&12\catcode `\#12\catcode `\^12\catcode `\_12\catcode `\%12\relax}%
	\providecommand \@@startlink[1]{}%
	\providecommand \@@endlink[0]{}%
	\providecommand \url  [0]{\begingroup\@sanitize@url \@url }%
	\providecommand \@url [1]{\endgroup\@href {#1}{\urlprefix }}%
	\providecommand \urlprefix  [0]{URL }%
	\providecommand \Eprint [0]{\href }%
	\providecommand \doibase [0]{https://doi.org/}%
	\providecommand \selectlanguage [0]{\@gobble}%
	\providecommand \bibinfo  [0]{\@secondoftwo}%
	\providecommand \bibfield  [0]{\@secondoftwo}%
	\providecommand \translation [1]{[#1]}%
	\providecommand \BibitemOpen [0]{}%
	\providecommand \bibitemStop [0]{}%
	\providecommand \bibitemNoStop [0]{.\EOS\space}%
	\providecommand \EOS [0]{\spacefactor3000\relax}%
	\providecommand \BibitemShut  [1]{\csname bibitem#1\endcsname}%
	\let\auto@bib@innerbib\@empty
	\bibitem [{\citenamefont {Awschalom}\ \emph {et~al.}(2018)\citenamefont
		{Awschalom}, \citenamefont {Hanson}, \citenamefont {Wrachtrup},\ and\
		\citenamefont {Zhou}}]{Awschalom2018}%
	\BibitemOpen
	\bibfield  {author} {\bibinfo {author} {\bibfnamefont {D.~D.}\ \bibnamefont
			{Awschalom}}, \bibinfo {author} {\bibfnamefont {R.}~\bibnamefont {Hanson}},
		\bibinfo {author} {\bibfnamefont {J.}~\bibnamefont {Wrachtrup}},\ and\
		\bibinfo {author} {\bibfnamefont {B.~B.}\ \bibnamefont {Zhou}},\ }\bibfield
	{title} {\bibinfo {title} {Quantum technologies with optically interfaced
			solid-state spins},\ }\href {https://doi.org/10.1038/s41566-018-0232-2}
	{\bibfield  {journal} {\bibinfo  {journal} {Nat. Photonics}\ }\textbf
		{\bibinfo {volume} {12}},\ \bibinfo {pages} {516} (\bibinfo {year}
		{2018})}\BibitemShut {NoStop}%
	\bibitem [{\citenamefont {Kalb}\ \emph {et~al.}(2017)\citenamefont {Kalb},
		\citenamefont {Reiserer}, \citenamefont {Humphreys}, \citenamefont
		{Bakermans}, \citenamefont {Kamerling}, \citenamefont {Nickerson},
		\citenamefont {Benjamin}, \citenamefont {Twitchen}, \citenamefont {Markham},\
		and\ \citenamefont {Hanson}}]{Kalb2017}%
	\BibitemOpen
	\bibfield  {author} {\bibinfo {author} {\bibfnamefont {N.}~\bibnamefont
			{Kalb}}, \bibinfo {author} {\bibfnamefont {A.~A.}\ \bibnamefont {Reiserer}},
		\bibinfo {author} {\bibfnamefont {P.~C.}\ \bibnamefont {Humphreys}}, \bibinfo
		{author} {\bibfnamefont {J.~J.~W.}\ \bibnamefont {Bakermans}}, \bibinfo
		{author} {\bibfnamefont {S.~J.}\ \bibnamefont {Kamerling}}, \bibinfo {author}
		{\bibfnamefont {N.~H.}\ \bibnamefont {Nickerson}}, \bibinfo {author}
		{\bibfnamefont {S.~C.}\ \bibnamefont {Benjamin}}, \bibinfo {author}
		{\bibfnamefont {D.~J.}\ \bibnamefont {Twitchen}}, \bibinfo {author}
		{\bibfnamefont {M.}~\bibnamefont {Markham}},\ and\ \bibinfo {author}
		{\bibfnamefont {R.}~\bibnamefont {Hanson}},\ }\bibfield  {title} {\bibinfo
		{title} {Entanglement distillation between solid-state quantum network
			nodes},\ }\href {https://doi.org/10.1126/science.aan0070} {\bibfield
		{journal} {\bibinfo  {journal} {Science}\ }\textbf {\bibinfo {volume}
			{356}},\ \bibinfo {pages} {928} (\bibinfo {year} {2017})}\BibitemShut
	{NoStop}%
	\bibitem [{\citenamefont {Humphreys}\ \emph {et~al.}(2018)\citenamefont
		{Humphreys}, \citenamefont {Kalb}, \citenamefont {Morits}, \citenamefont
		{Schouten}, \citenamefont {Vermeulen}, \citenamefont {Twitchen},
		\citenamefont {Markham},\ and\ \citenamefont {Hanson}}]{Humphreys2018}%
	\BibitemOpen
	\bibfield  {author} {\bibinfo {author} {\bibfnamefont {P.~C.}\ \bibnamefont
			{Humphreys}}, \bibinfo {author} {\bibfnamefont {N.}~\bibnamefont {Kalb}},
		\bibinfo {author} {\bibfnamefont {J.~P.~J.}\ \bibnamefont {Morits}}, \bibinfo
		{author} {\bibfnamefont {R.~N.}\ \bibnamefont {Schouten}}, \bibinfo {author}
		{\bibfnamefont {R.~F.~L.}\ \bibnamefont {Vermeulen}}, \bibinfo {author}
		{\bibfnamefont {D.~J.}\ \bibnamefont {Twitchen}}, \bibinfo {author}
		{\bibfnamefont {M.}~\bibnamefont {Markham}},\ and\ \bibinfo {author}
		{\bibfnamefont {R.}~\bibnamefont {Hanson}},\ }\bibfield  {title} {\bibinfo
		{title} {Deterministic delivery of remote entanglement on a quantum
			network},\ }\href {https://doi.org/10.1038/s41586-018-0200-5} {\bibfield
		{journal} {\bibinfo  {journal} {Nature}\ }\textbf {\bibinfo {volume} {558}},\
		\bibinfo {pages} {268} (\bibinfo {year} {2018})}\BibitemShut {NoStop}%
	\bibitem [{\citenamefont {Janitz}\ \emph {et~al.}(2020)\citenamefont {Janitz},
		\citenamefont {Bhaskar},\ and\ \citenamefont {Childress}}]{Janitz2020}%
	\BibitemOpen
	\bibfield  {author} {\bibinfo {author} {\bibfnamefont {E.}~\bibnamefont
			{Janitz}}, \bibinfo {author} {\bibfnamefont {M.~K.}\ \bibnamefont
			{Bhaskar}},\ and\ \bibinfo {author} {\bibfnamefont {L.}~\bibnamefont
			{Childress}},\ }\bibfield  {title} {\bibinfo {title} {Cavity quantum
			electrodynamics with color centers in diamond},\ }\href
	{https://doi.org/10.1364/OPTICA.398628} {\bibfield  {journal} {\bibinfo
			{journal} {Optica}\ }\textbf {\bibinfo {volume} {7}},\ \bibinfo {pages}
		{1232} (\bibinfo {year} {2020})}\BibitemShut {NoStop}%
	\bibitem [{\citenamefont {Lukin}\ \emph
		{et~al.}(2020{\natexlab{a}})\citenamefont {Lukin}, \citenamefont {Guidry},\
		and\ \citenamefont {Vu\ifmmode \check{c}\else
			\v{c}\fi{}kovi\ifmmode~\acute{c}\else \'{c}\fi{}}}]{Daniil2020}%
	\BibitemOpen
	\bibfield  {author} {\bibinfo {author} {\bibfnamefont {D.~M.}\ \bibnamefont
			{Lukin}}, \bibinfo {author} {\bibfnamefont {M.~A.}\ \bibnamefont {Guidry}},\
		and\ \bibinfo {author} {\bibfnamefont {J.}~\bibnamefont {Vu\ifmmode
				\check{c}\else \v{c}\fi{}kovi\ifmmode~\acute{c}\else \'{c}\fi{}}},\
	}\bibfield  {title} {\bibinfo {title} {Integrated quantum photonics with
			silicon carbide: Challenges and prospects},\ }\href
	{https://doi.org/10.1103/PRXQuantum.1.020102} {\bibfield  {journal} {\bibinfo
			{journal} {PRX Quantum}\ }\textbf {\bibinfo {volume} {1}},\ \bibinfo {pages}
		{020102} (\bibinfo {year} {2020}{\natexlab{a}})}\BibitemShut {NoStop}%
	\bibitem [{\citenamefont {Borregaard}\ \emph {et~al.}(2019)\citenamefont
		{Borregaard}, \citenamefont {Sørensen},\ and\ \citenamefont
		{Lodahl}}]{Borregaard2019}%
	\BibitemOpen
	\bibfield  {author} {\bibinfo {author} {\bibfnamefont {J.}~\bibnamefont
			{Borregaard}}, \bibinfo {author} {\bibfnamefont {A.~S.}\ \bibnamefont
			{Sørensen}},\ and\ \bibinfo {author} {\bibfnamefont {P.}~\bibnamefont
			{Lodahl}},\ }\bibfield  {title} {\bibinfo {title} {Quantum networks with
			deterministic spin–photon interfaces},\ }\href
	{https://doi.org/https://doi.org/10.1002/qute.201800091} {\bibfield
		{journal} {\bibinfo  {journal} {Adv. Quantum Technol.}\ }\textbf {\bibinfo
			{volume} {2}},\ \bibinfo {pages} {1800091} (\bibinfo {year}
		{2019})}\BibitemShut {NoStop}%
	\bibitem [{\citenamefont {Degen}\ \emph {et~al.}(2017)\citenamefont {Degen},
		\citenamefont {Reinhard},\ and\ \citenamefont {Cappellaro}}]{Degen2017}%
	\BibitemOpen
	\bibfield  {author} {\bibinfo {author} {\bibfnamefont {C.~L.}\ \bibnamefont
			{Degen}}, \bibinfo {author} {\bibfnamefont {F.}~\bibnamefont {Reinhard}},\
		and\ \bibinfo {author} {\bibfnamefont {P.}~\bibnamefont {Cappellaro}},\
	}\bibfield  {title} {\bibinfo {title} {Quantum sensing},\ }\href
	{https://doi.org/10.1103/RevModPhys.89.035002} {\bibfield  {journal}
		{\bibinfo  {journal} {Rev. Mod. Phys.}\ }\textbf {\bibinfo {volume} {89}},\
		\bibinfo {pages} {035002} (\bibinfo {year} {2017})}\BibitemShut {NoStop}%
	\bibitem [{\citenamefont {Widmann}\ \emph {et~al.}(2015)\citenamefont
		{Widmann}, \citenamefont {Lee}, \citenamefont {Rendler}, \citenamefont {Son},
		\citenamefont {Fedder}, \citenamefont {Paik}, \citenamefont {Yang},
		\citenamefont {Zhao}, \citenamefont {Yang}, \citenamefont {Booker},
		\citenamefont {Denisenko}, \citenamefont {Jamali}, \citenamefont
		{Momenzadeh}, \citenamefont {Gerhardt}, \citenamefont {Ohshima},
		\citenamefont {Gali}, \citenamefont {Janz{\'e}n},\ and\ \citenamefont
		{Wrachtrup}}]{Widmann2015}%
	\BibitemOpen
	\bibfield  {author} {\bibinfo {author} {\bibfnamefont {M.}~\bibnamefont
			{Widmann}}, \bibinfo {author} {\bibfnamefont {S.-Y.}\ \bibnamefont {Lee}},
		\bibinfo {author} {\bibfnamefont {T.}~\bibnamefont {Rendler}}, \bibinfo
		{author} {\bibfnamefont {N.~T.}\ \bibnamefont {Son}}, \bibinfo {author}
		{\bibfnamefont {H.}~\bibnamefont {Fedder}}, \bibinfo {author} {\bibfnamefont
			{S.}~\bibnamefont {Paik}}, \bibinfo {author} {\bibfnamefont {L.-P.}\
			\bibnamefont {Yang}}, \bibinfo {author} {\bibfnamefont {N.}~\bibnamefont
			{Zhao}}, \bibinfo {author} {\bibfnamefont {S.}~\bibnamefont {Yang}}, \bibinfo
		{author} {\bibfnamefont {I.}~\bibnamefont {Booker}}, \bibinfo {author}
		{\bibfnamefont {A.}~\bibnamefont {Denisenko}}, \bibinfo {author}
		{\bibfnamefont {M.}~\bibnamefont {Jamali}}, \bibinfo {author} {\bibfnamefont
			{S.~A.}\ \bibnamefont {Momenzadeh}}, \bibinfo {author} {\bibfnamefont
			{I.}~\bibnamefont {Gerhardt}}, \bibinfo {author} {\bibfnamefont
			{T.}~\bibnamefont {Ohshima}}, \bibinfo {author} {\bibfnamefont
			{A.}~\bibnamefont {Gali}}, \bibinfo {author} {\bibfnamefont {E.}~\bibnamefont
			{Janz{\'e}n}},\ and\ \bibinfo {author} {\bibfnamefont {J.}~\bibnamefont
			{Wrachtrup}},\ }\bibfield  {title} {\bibinfo {title} {Coherent control of
			single spins in silicon carbide at room temperature},\ }\href
	{https://doi.org/10.1038/nmat4145} {\bibfield  {journal} {\bibinfo  {journal}
			{Nat. Mater.}\ }\textbf {\bibinfo {volume} {14}},\ \bibinfo {pages} {164}
		(\bibinfo {year} {2015})}\BibitemShut {NoStop}%
	\bibitem [{\citenamefont {Neumann}\ \emph {et~al.}(2009)\citenamefont
		{Neumann}, \citenamefont {Kolesov}, \citenamefont {Jacques}, \citenamefont
		{Beck}, \citenamefont {Tisler}, \citenamefont {Batalov}, \citenamefont
		{Rogers}, \citenamefont {Manson}, \citenamefont {Balasubramanian},
		\citenamefont {Jelezko},\ and\ \citenamefont {Wrachtrup}}]{Neumann2009}%
	\BibitemOpen
	\bibfield  {author} {\bibinfo {author} {\bibfnamefont {P.}~\bibnamefont
			{Neumann}}, \bibinfo {author} {\bibfnamefont {R.}~\bibnamefont {Kolesov}},
		\bibinfo {author} {\bibfnamefont {V.}~\bibnamefont {Jacques}}, \bibinfo
		{author} {\bibfnamefont {J.}~\bibnamefont {Beck}}, \bibinfo {author}
		{\bibfnamefont {J.}~\bibnamefont {Tisler}}, \bibinfo {author} {\bibfnamefont
			{A.}~\bibnamefont {Batalov}}, \bibinfo {author} {\bibfnamefont
			{L.}~\bibnamefont {Rogers}}, \bibinfo {author} {\bibfnamefont {N.~B.}\
			\bibnamefont {Manson}}, \bibinfo {author} {\bibfnamefont {G.}~\bibnamefont
			{Balasubramanian}}, \bibinfo {author} {\bibfnamefont {F.}~\bibnamefont
			{Jelezko}},\ and\ \bibinfo {author} {\bibfnamefont {J.}~\bibnamefont
			{Wrachtrup}},\ }\bibfield  {title} {\bibinfo {title} {Excited-state
			spectroscopy of single {NV} defects in diamond using optically detected
			magnetic resonance},\ }\href {https://doi.org/10.1088/1367-2630/11/1/013017}
	{\bibfield  {journal} {\bibinfo  {journal} {New J. Phys.}\ }\textbf {\bibinfo
			{volume} {11}},\ \bibinfo {pages} {013017} (\bibinfo {year}
		{2009})}\BibitemShut {NoStop}%
	\bibitem [{\citenamefont {Singh}\ \emph {et~al.}(2020)\citenamefont {Singh},
		\citenamefont {Anisimov}, \citenamefont {Nagalyuk}, \citenamefont {Mokhov},
		\citenamefont {Baranov},\ and\ \citenamefont {Suter}}]{Singh2020}%
	\BibitemOpen
	\bibfield  {author} {\bibinfo {author} {\bibfnamefont {H.}~\bibnamefont
			{Singh}}, \bibinfo {author} {\bibfnamefont {A.~N.}\ \bibnamefont {Anisimov}},
		\bibinfo {author} {\bibfnamefont {S.~S.}\ \bibnamefont {Nagalyuk}}, \bibinfo
		{author} {\bibfnamefont {E.~N.}\ \bibnamefont {Mokhov}}, \bibinfo {author}
		{\bibfnamefont {P.~G.}\ \bibnamefont {Baranov}},\ and\ \bibinfo {author}
		{\bibfnamefont {D.}~\bibnamefont {Suter}},\ }\bibfield  {title} {\bibinfo
		{title} {Experimental characterization of spin-$\frac{3}{2}$ silicon vacancy
			centers in $6H$-SiC},\ }\href {https://doi.org/10.1103/PhysRevB.101.134110}
	{\bibfield  {journal} {\bibinfo  {journal} {Phys. Rev. B}\ }\textbf {\bibinfo
			{volume} {101}},\ \bibinfo {pages} {134110} (\bibinfo {year}
		{2020})}\BibitemShut {NoStop}%
	\bibitem [{\citenamefont {Jelezko}\ \emph {et~al.}(2004)\citenamefont
		{Jelezko}, \citenamefont {Gaebel}, \citenamefont {Popa}, \citenamefont
		{Gruber},\ and\ \citenamefont {Wrachtrup}}]{Jelezko2004}%
	\BibitemOpen
	\bibfield  {author} {\bibinfo {author} {\bibfnamefont {F.}~\bibnamefont
			{Jelezko}}, \bibinfo {author} {\bibfnamefont {T.}~\bibnamefont {Gaebel}},
		\bibinfo {author} {\bibfnamefont {I.}~\bibnamefont {Popa}}, \bibinfo {author}
		{\bibfnamefont {A.}~\bibnamefont {Gruber}},\ and\ \bibinfo {author}
		{\bibfnamefont {J.}~\bibnamefont {Wrachtrup}},\ }\bibfield  {title} {\bibinfo
		{title} {Observation of coherent oscillations in a single electron spin},\
	}\href {https://doi.org/10.1103/PhysRevLett.92.076401} {\bibfield  {journal}
		{\bibinfo  {journal} {Phys. Rev. Lett.}\ }\textbf {\bibinfo {volume} {92}},\
		\bibinfo {pages} {076401} (\bibinfo {year} {2004})}\BibitemShut {NoStop}%
	\bibitem [{\citenamefont {Falk}\ \emph {et~al.}(2013)\citenamefont {Falk},
		\citenamefont {Buckley}, \citenamefont {Calusine}, \citenamefont {Koehl},
		\citenamefont {Dobrovitski}, \citenamefont {Politi}, \citenamefont {Zorman},
		\citenamefont {Feng},\ and\ \citenamefont {Awschalom}}]{Falk2013}%
	\BibitemOpen
	\bibfield  {author} {\bibinfo {author} {\bibfnamefont {A.~L.}\ \bibnamefont
			{Falk}}, \bibinfo {author} {\bibfnamefont {B.~B.}\ \bibnamefont {Buckley}},
		\bibinfo {author} {\bibfnamefont {G.}~\bibnamefont {Calusine}}, \bibinfo
		{author} {\bibfnamefont {W.~F.}\ \bibnamefont {Koehl}}, \bibinfo {author}
		{\bibfnamefont {V.~V.}\ \bibnamefont {Dobrovitski}}, \bibinfo {author}
		{\bibfnamefont {A.}~\bibnamefont {Politi}}, \bibinfo {author} {\bibfnamefont
			{C.~A.}\ \bibnamefont {Zorman}}, \bibinfo {author} {\bibfnamefont {P.~X.-L.}\
			\bibnamefont {Feng}},\ and\ \bibinfo {author} {\bibfnamefont {D.~D.}\
			\bibnamefont {Awschalom}},\ }\bibfield  {title} {\bibinfo {title} {Polytype
			control of spin qubits in silicon carbide},\ }\href
	{https://doi.org/10.1038/ncomms2854} {\bibfield  {journal} {\bibinfo
			{journal} {Nat. Commun.}\ }\textbf {\bibinfo {volume} {4}},\ \bibinfo {pages}
		{1819} (\bibinfo {year} {2013})}\BibitemShut {NoStop}%
	\bibitem [{\citenamefont {Song}\ \emph {et~al.}(2020)\citenamefont {Song},
		\citenamefont {Tian}, \citenamefont {Hu}, \citenamefont {Zhou}, \citenamefont
		{Xing}, \citenamefont {Lu}, \citenamefont {Chen}, \citenamefont {Wang},
		\citenamefont {Xu},\ and\ \citenamefont {Du}}]{Song2020}%
	\BibitemOpen
	\bibfield  {author} {\bibinfo {author} {\bibfnamefont {Y.}~\bibnamefont
			{Song}}, \bibinfo {author} {\bibfnamefont {Y.}~\bibnamefont {Tian}}, \bibinfo
		{author} {\bibfnamefont {Z.}~\bibnamefont {Hu}}, \bibinfo {author}
		{\bibfnamefont {F.}~\bibnamefont {Zhou}}, \bibinfo {author} {\bibfnamefont
			{T.}~\bibnamefont {Xing}}, \bibinfo {author} {\bibfnamefont {D.}~\bibnamefont
			{Lu}}, \bibinfo {author} {\bibfnamefont {B.}~\bibnamefont {Chen}}, \bibinfo
		{author} {\bibfnamefont {Y.}~\bibnamefont {Wang}}, \bibinfo {author}
		{\bibfnamefont {N.}~\bibnamefont {Xu}},\ and\ \bibinfo {author}
		{\bibfnamefont {J.}~\bibnamefont {Du}},\ }\bibfield  {title} {\bibinfo
		{title} {Pulse-width-induced polarization enhancement of optically pumped n-v
			electron spin in diamond},\ }\href {https://doi.org/10.1364/PRJ.386983}
	{\bibfield  {journal} {\bibinfo  {journal} {Photon. Res.}\ }\textbf {\bibinfo
			{volume} {8}},\ \bibinfo {pages} {1289} (\bibinfo {year} {2020})}\BibitemShut
	{NoStop}%
	\bibitem [{\citenamefont {Nagy}\ \emph {et~al.}(2019)\citenamefont {Nagy},
		\citenamefont {Niethammer}, \citenamefont {Widmann}, \citenamefont {Chen},
		\citenamefont {Udvarhelyi}, \citenamefont {Bonato}, \citenamefont {Hassan},
		\citenamefont {Karhu}, \citenamefont {Ivanov}, \citenamefont {Son},
		\citenamefont {Maze}, \citenamefont {Ohshima}, \citenamefont {Soykal},
		\citenamefont {Gali}, \citenamefont {Lee}, \citenamefont {Kaiser},\ and\
		\citenamefont {Wrachtrup}}]{Nagy2019}%
	\BibitemOpen
	\bibfield  {author} {\bibinfo {author} {\bibfnamefont {R.}~\bibnamefont
			{Nagy}}, \bibinfo {author} {\bibfnamefont {M.}~\bibnamefont {Niethammer}},
		\bibinfo {author} {\bibfnamefont {M.}~\bibnamefont {Widmann}}, \bibinfo
		{author} {\bibfnamefont {Y.-C.}\ \bibnamefont {Chen}}, \bibinfo {author}
		{\bibfnamefont {P.}~\bibnamefont {Udvarhelyi}}, \bibinfo {author}
		{\bibfnamefont {C.}~\bibnamefont {Bonato}}, \bibinfo {author} {\bibfnamefont
			{J.~U.}\ \bibnamefont {Hassan}}, \bibinfo {author} {\bibfnamefont
			{R.}~\bibnamefont {Karhu}}, \bibinfo {author} {\bibfnamefont {I.~G.}\
			\bibnamefont {Ivanov}}, \bibinfo {author} {\bibfnamefont {N.~T.}\
			\bibnamefont {Son}}, \bibinfo {author} {\bibfnamefont {J.~R.}\ \bibnamefont
			{Maze}}, \bibinfo {author} {\bibfnamefont {T.}~\bibnamefont {Ohshima}},
		\bibinfo {author} {\bibfnamefont {{\"O}.~O.}\ \bibnamefont {Soykal}},
		\bibinfo {author} {\bibfnamefont {{\'A}.}~\bibnamefont {Gali}}, \bibinfo
		{author} {\bibfnamefont {S.-Y.}\ \bibnamefont {Lee}}, \bibinfo {author}
		{\bibfnamefont {F.}~\bibnamefont {Kaiser}},\ and\ \bibinfo {author}
		{\bibfnamefont {J.}~\bibnamefont {Wrachtrup}},\ }\bibfield  {title} {\bibinfo
		{title} {High-fidelity spin and optical control of single silicon-vacancy
			centres in silicon carbide},\ }\href
	{https://doi.org/10.1038/s41467-019-09873-9} {\bibfield  {journal} {\bibinfo
			{journal} {Nat. Commun.}\ }\textbf {\bibinfo {volume} {10}},\ \bibinfo
		{pages} {1954} (\bibinfo {year} {2019})}\BibitemShut {NoStop}%
	\bibitem [{\citenamefont {Morioka}\ \emph {et~al.}(2020)\citenamefont
		{Morioka}, \citenamefont {Babin}, \citenamefont {Nagy}, \citenamefont
		{Gediz}, \citenamefont {Hesselmeier}, \citenamefont {Liu}, \citenamefont
		{Joliffe}, \citenamefont {Niethammer}, \citenamefont {Dasari}, \citenamefont
		{Vorobyov}, \citenamefont {Kolesov}, \citenamefont {St{\"o}hr}, \citenamefont
		{Ul-Hassan}, \citenamefont {Son}, \citenamefont {Ohshima}, \citenamefont
		{Udvarhelyi}, \citenamefont {Thiering}, \citenamefont {Gali}, \citenamefont
		{Wrachtrup},\ and\ \citenamefont {Kaiser}}]{Morioka2020}%
	\BibitemOpen
	\bibfield  {author} {\bibinfo {author} {\bibfnamefont {N.}~\bibnamefont
			{Morioka}}, \bibinfo {author} {\bibfnamefont {C.}~\bibnamefont {Babin}},
		\bibinfo {author} {\bibfnamefont {R.}~\bibnamefont {Nagy}}, \bibinfo {author}
		{\bibfnamefont {I.}~\bibnamefont {Gediz}}, \bibinfo {author} {\bibfnamefont
			{E.}~\bibnamefont {Hesselmeier}}, \bibinfo {author} {\bibfnamefont
			{D.}~\bibnamefont {Liu}}, \bibinfo {author} {\bibfnamefont {M.}~\bibnamefont
			{Joliffe}}, \bibinfo {author} {\bibfnamefont {M.}~\bibnamefont {Niethammer}},
		\bibinfo {author} {\bibfnamefont {D.}~\bibnamefont {Dasari}}, \bibinfo
		{author} {\bibfnamefont {V.}~\bibnamefont {Vorobyov}}, \bibinfo {author}
		{\bibfnamefont {R.}~\bibnamefont {Kolesov}}, \bibinfo {author} {\bibfnamefont
			{R.}~\bibnamefont {St{\"o}hr}}, \bibinfo {author} {\bibfnamefont
			{J.}~\bibnamefont {Ul-Hassan}}, \bibinfo {author} {\bibfnamefont {N.~T.}\
			\bibnamefont {Son}}, \bibinfo {author} {\bibfnamefont {T.}~\bibnamefont
			{Ohshima}}, \bibinfo {author} {\bibfnamefont {P.}~\bibnamefont {Udvarhelyi}},
		\bibinfo {author} {\bibfnamefont {G.}~\bibnamefont {Thiering}}, \bibinfo
		{author} {\bibfnamefont {A.}~\bibnamefont {Gali}}, \bibinfo {author}
		{\bibfnamefont {J.}~\bibnamefont {Wrachtrup}},\ and\ \bibinfo {author}
		{\bibfnamefont {F.}~\bibnamefont {Kaiser}},\ }\bibfield  {title} {\bibinfo
		{title} {Spin-controlled generation of indistinguishable and distinguishable
			photons from silicon vacancy centres in silicon carbide},\ }\href
	{https://doi.org/10.1038/s41467-020-16330-5} {\bibfield  {journal} {\bibinfo
			{journal} {Nat. Commun.}\ }\textbf {\bibinfo {volume} {11}},\ \bibinfo
		{pages} {2516} (\bibinfo {year} {2020})}\BibitemShut {NoStop}%
	\bibitem [{\citenamefont {Udvarhelyi}\ \emph {et~al.}(2020)\citenamefont
		{Udvarhelyi}, \citenamefont {Thiering}, \citenamefont {Morioka},
		\citenamefont {Babin}, \citenamefont {Kaiser}, \citenamefont {Lukin},
		\citenamefont {Ohshima}, \citenamefont {Ul-Hassan}, \citenamefont {Son},
		\citenamefont {Vu\ifmmode \check{c}\else
			\v{c}\fi{}kovi\ifmmode~\acute{c}\else \'{c}\fi{}}, \citenamefont
		{Wrachtrup},\ and\ \citenamefont {Gali}}]{Gali2020}%
	\BibitemOpen
	\bibfield  {author} {\bibinfo {author} {\bibfnamefont {P.}~\bibnamefont
			{Udvarhelyi}}, \bibinfo {author} {\bibfnamefont {G.~m.~H.}\ \bibnamefont
			{Thiering}}, \bibinfo {author} {\bibfnamefont {N.}~\bibnamefont {Morioka}},
		\bibinfo {author} {\bibfnamefont {C.}~\bibnamefont {Babin}}, \bibinfo
		{author} {\bibfnamefont {F.}~\bibnamefont {Kaiser}}, \bibinfo {author}
		{\bibfnamefont {D.}~\bibnamefont {Lukin}}, \bibinfo {author} {\bibfnamefont
			{T.}~\bibnamefont {Ohshima}}, \bibinfo {author} {\bibfnamefont
			{J.}~\bibnamefont {Ul-Hassan}}, \bibinfo {author} {\bibfnamefont {N.~T.}\
			\bibnamefont {Son}}, \bibinfo {author} {\bibfnamefont {J.}~\bibnamefont
			{Vu\ifmmode \check{c}\else \v{c}\fi{}kovi\ifmmode~\acute{c}\else
				\'{c}\fi{}}}, \bibinfo {author} {\bibfnamefont {J.}~\bibnamefont
			{Wrachtrup}},\ and\ \bibinfo {author} {\bibfnamefont {A.}~\bibnamefont
			{Gali}},\ }\bibfield  {title} {\bibinfo {title} {Vibronic states and their
			effect on the temperature and strain dependence of silicon-vacancy qubits in
			$4H$-$\mathrm{Si}\mathrm{C}$},\ }\href
	{https://doi.org/10.1103/PhysRevApplied.13.054017} {\bibfield  {journal}
		{\bibinfo  {journal} {Phys. Rev. Applied}\ }\textbf {\bibinfo {volume}
			{13}},\ \bibinfo {pages} {054017} (\bibinfo {year} {2020})}\BibitemShut
	{NoStop}%
	\bibitem [{\citenamefont {Robledo}\ \emph
		{et~al.}(2011{\natexlab{a}})\citenamefont {Robledo}, \citenamefont {Bernien},
		\citenamefont {van~der Sar},\ and\ \citenamefont {Hanson}}]{Robledo2011njp}%
	\BibitemOpen
	\bibfield  {author} {\bibinfo {author} {\bibfnamefont {L.}~\bibnamefont
			{Robledo}}, \bibinfo {author} {\bibfnamefont {H.}~\bibnamefont {Bernien}},
		\bibinfo {author} {\bibfnamefont {T.}~\bibnamefont {van~der Sar}},\ and\
		\bibinfo {author} {\bibfnamefont {R.}~\bibnamefont {Hanson}},\ }\bibfield
	{title} {\bibinfo {title} {Spin dynamics in the optical cycle of single
			nitrogen-vacancy centres in diamond},\ }\href
	{https://doi.org/10.1088/1367-2630/13/2/025013} {\bibfield  {journal}
		{\bibinfo  {journal} {New J. Phys.}\ }\textbf {\bibinfo {volume} {13}},\
		\bibinfo {pages} {025013} (\bibinfo {year} {2011}{\natexlab{a}})}\BibitemShut
	{NoStop}%
	\bibitem [{\citenamefont {Manson}\ \emph {et~al.}(2006)\citenamefont {Manson},
		\citenamefont {Harrison},\ and\ \citenamefont {Sellars}}]{Manson2006}%
	\BibitemOpen
	\bibfield  {author} {\bibinfo {author} {\bibfnamefont {N.~B.}\ \bibnamefont
			{Manson}}, \bibinfo {author} {\bibfnamefont {J.~P.}\ \bibnamefont
			{Harrison}},\ and\ \bibinfo {author} {\bibfnamefont {M.~J.}\ \bibnamefont
			{Sellars}},\ }\bibfield  {title} {\bibinfo {title} {Nitrogen-vacancy center
			in diamond: Model of the electronic structure and associated dynamics},\
	}\href {https://doi.org/10.1103/PhysRevB.74.104303} {\bibfield  {journal}
		{\bibinfo  {journal} {Phys. Rev. B}\ }\textbf {\bibinfo {volume} {74}},\
		\bibinfo {pages} {104303} (\bibinfo {year} {2006})}\BibitemShut {NoStop}%
	\bibitem [{\citenamefont {Fuchs}\ \emph {et~al.}(2015)\citenamefont {Fuchs},
		\citenamefont {Stender}, \citenamefont {Trupke}, \citenamefont {Simin},
		\citenamefont {Pflaum}, \citenamefont {Dyakonov},\ and\ \citenamefont
		{Astakhov}}]{Fuchs2015}%
	\BibitemOpen
	\bibfield  {author} {\bibinfo {author} {\bibfnamefont {F.}~\bibnamefont
			{Fuchs}}, \bibinfo {author} {\bibfnamefont {B.}~\bibnamefont {Stender}},
		\bibinfo {author} {\bibfnamefont {M.}~\bibnamefont {Trupke}}, \bibinfo
		{author} {\bibfnamefont {D.}~\bibnamefont {Simin}}, \bibinfo {author}
		{\bibfnamefont {J.}~\bibnamefont {Pflaum}}, \bibinfo {author} {\bibfnamefont
			{V.}~\bibnamefont {Dyakonov}},\ and\ \bibinfo {author} {\bibfnamefont
			{G.~V.}\ \bibnamefont {Astakhov}},\ }\bibfield  {title} {\bibinfo {title}
		{Engineering near-infrared single-photon emitters with optically active spins
			in ultrapure silicon carbide},\ }\href {https://doi.org/10.1038/ncomms8578}
	{\bibfield  {journal} {\bibinfo  {journal} {Nat. Commun.}\ }\textbf {\bibinfo
			{volume} {6}},\ \bibinfo {pages} {7578} (\bibinfo {year} {2015})}\BibitemShut
	{NoStop}%
	\bibitem [{\citenamefont {Neu}\ \emph {et~al.}(2012)\citenamefont {Neu},
		\citenamefont {Agio},\ and\ \citenamefont {Christoph}}]{Elke2012}%
	\BibitemOpen
	\bibfield  {author} {\bibinfo {author} {\bibfnamefont {E.}~\bibnamefont
			{Neu}}, \bibinfo {author} {\bibfnamefont {M.}~\bibnamefont {Agio}},\ and\
		\bibinfo {author} {\bibnamefont {Christoph}},\ }\bibfield  {title} {\bibinfo
		{title} {Photophysics of single silicon vacancy centers in diamond:
			implications for single photon emission},\ }\href
	{https://doi.org/10.1364/OE.20.019956} {\bibfield  {journal} {\bibinfo
			{journal} {Opt. Express}\ }\textbf {\bibinfo {volume} {20}},\ \bibinfo
		{pages} {19956} (\bibinfo {year} {2012})}\BibitemShut {NoStop}%
	\bibitem [{\citenamefont {Castelletto}\ \emph {et~al.}(2015)\citenamefont
		{Castelletto}, \citenamefont {Rosa},\ and\ \citenamefont
		{Johnson}}]{Castelletto2015}%
	\BibitemOpen
	\bibfield  {author} {\bibinfo {author} {\bibfnamefont {S.}~\bibnamefont
			{Castelletto}}, \bibinfo {author} {\bibfnamefont {L.}~\bibnamefont {Rosa}},\
		and\ \bibinfo {author} {\bibfnamefont {B.~C.}\ \bibnamefont {Johnson}},\
	}\bibinfo {title} {Advanced silicon carbide devicesa and processing}\
	(\bibinfo  {publisher} {IntechOpen},\ \bibinfo {year} {2015})\ Chap.\
	\bibinfo {chapter} {Silicon carbide for novel quantum technology
		devices}\BibitemShut {NoStop}%
	\bibitem [{\citenamefont {McCloskey}\ \emph {et~al.}(2014)\citenamefont
		{McCloskey}, \citenamefont {Fox}, \citenamefont {O'Hara}, \citenamefont
		{Usov}, \citenamefont {Scanlan}, \citenamefont {McEvoy}, \citenamefont
		{Duesberg}, \citenamefont {Cross}, \citenamefont {Zhang},\ and\ \citenamefont
		{Donegan}}]{McCloskey2014}%
	\BibitemOpen
	\bibfield  {author} {\bibinfo {author} {\bibfnamefont {D.}~\bibnamefont
			{McCloskey}}, \bibinfo {author} {\bibfnamefont {D.}~\bibnamefont {Fox}},
		\bibinfo {author} {\bibfnamefont {N.}~\bibnamefont {O'Hara}}, \bibinfo
		{author} {\bibfnamefont {V.}~\bibnamefont {Usov}}, \bibinfo {author}
		{\bibfnamefont {D.}~\bibnamefont {Scanlan}}, \bibinfo {author} {\bibfnamefont
			{N.}~\bibnamefont {McEvoy}}, \bibinfo {author} {\bibfnamefont {G.~S.}\
			\bibnamefont {Duesberg}}, \bibinfo {author} {\bibfnamefont {G.~L.~W.}\
			\bibnamefont {Cross}}, \bibinfo {author} {\bibfnamefont {H.~Z.}\ \bibnamefont
			{Zhang}},\ and\ \bibinfo {author} {\bibfnamefont {J.~F.}\ \bibnamefont
			{Donegan}},\ }\bibfield  {title} {\bibinfo {title} {Helium ion microscope
			generated nitrogen-vacancy centres in type ib diamond},\ }\href
	{https://doi.org/10.1063/1.4862331} {\bibfield  {journal} {\bibinfo
			{journal} {Appl. Phys. Lett.}\ }\textbf {\bibinfo {volume} {104}},\ \bibinfo
		{pages} {031109} (\bibinfo {year} {2014})}\BibitemShut {NoStop}%
	\bibitem [{\citenamefont {Radko}\ \emph {et~al.}(2016)\citenamefont {Radko},
		\citenamefont {Boll}, \citenamefont {Israelsen}, \citenamefont {Raatz},
		\citenamefont {Meijer}, \citenamefont {Jelezko}, \citenamefont {Andersen},\
		and\ \citenamefont {Huck}}]{Radko2016}%
	\BibitemOpen
	\bibfield  {author} {\bibinfo {author} {\bibfnamefont {I.~P.}\ \bibnamefont
			{Radko}}, \bibinfo {author} {\bibfnamefont {M.}~\bibnamefont {Boll}},
		\bibinfo {author} {\bibfnamefont {N.~M.}\ \bibnamefont {Israelsen}}, \bibinfo
		{author} {\bibfnamefont {N.}~\bibnamefont {Raatz}}, \bibinfo {author}
		{\bibfnamefont {J.}~\bibnamefont {Meijer}}, \bibinfo {author} {\bibfnamefont
			{F.}~\bibnamefont {Jelezko}}, \bibinfo {author} {\bibfnamefont {U.~L.}\
			\bibnamefont {Andersen}},\ and\ \bibinfo {author} {\bibfnamefont
			{A.}~\bibnamefont {Huck}},\ }\bibfield  {title} {\bibinfo {title}
		{Determining the internal quantum efficiency of shallow-implanted
			nitrogen-vacancy defects in bulk diamond},\ }\href
	{https://doi.org/10.1364/OE.24.027715} {\bibfield  {journal} {\bibinfo
			{journal} {Opt. Express}\ }\textbf {\bibinfo {volume} {24}},\ \bibinfo
		{pages} {27715} (\bibinfo {year} {2016})}\BibitemShut {NoStop}%
	\bibitem [{\citenamefont {Green}\ \emph {et~al.}(2017)\citenamefont {Green},
		\citenamefont {Mottishaw}, \citenamefont {Breeze}, \citenamefont {Edmonds},
		\citenamefont {D'Haenens-Johansson}, \citenamefont {Doherty}, \citenamefont
		{Williams}, \citenamefont {Twitchen},\ and\ \citenamefont
		{Newton}}]{Green2017}%
	\BibitemOpen
	\bibfield  {author} {\bibinfo {author} {\bibfnamefont {B.~L.}\ \bibnamefont
			{Green}}, \bibinfo {author} {\bibfnamefont {S.}~\bibnamefont {Mottishaw}},
		\bibinfo {author} {\bibfnamefont {B.~G.}\ \bibnamefont {Breeze}}, \bibinfo
		{author} {\bibfnamefont {A.~M.}\ \bibnamefont {Edmonds}}, \bibinfo {author}
		{\bibfnamefont {U.~F.~S.}\ \bibnamefont {D'Haenens-Johansson}}, \bibinfo
		{author} {\bibfnamefont {M.~W.}\ \bibnamefont {Doherty}}, \bibinfo {author}
		{\bibfnamefont {S.~D.}\ \bibnamefont {Williams}}, \bibinfo {author}
		{\bibfnamefont {D.~J.}\ \bibnamefont {Twitchen}},\ and\ \bibinfo {author}
		{\bibfnamefont {M.~E.}\ \bibnamefont {Newton}},\ }\bibfield  {title}
	{\bibinfo {title} {Neutral silicon-vacancy center in diamond: Spin
			polarization and lifetimes},\ }\href
	{https://doi.org/10.1103/PhysRevLett.119.096402} {\bibfield  {journal}
		{\bibinfo  {journal} {Phys. Rev. Lett.}\ }\textbf {\bibinfo {volume} {119}},\
		\bibinfo {pages} {096402} (\bibinfo {year} {2017})}\BibitemShut {NoStop}%
	\bibitem [{\citenamefont {Banks}\ \emph {et~al.}(2019)\citenamefont {Banks},
		\citenamefont {Soykal}, \citenamefont {Myers-Ward}, \citenamefont {Gaskill},
		\citenamefont {Reinecke},\ and\ \citenamefont {Carter}}]{Banks2019}%
	\BibitemOpen
	\bibfield  {author} {\bibinfo {author} {\bibfnamefont {H.~B.}\ \bibnamefont
			{Banks}}, \bibinfo {author} {\bibfnamefont {O.~O.}\ \bibnamefont {Soykal}},
		\bibinfo {author} {\bibfnamefont {R.~L.}\ \bibnamefont {Myers-Ward}},
		\bibinfo {author} {\bibfnamefont {D.~K.}\ \bibnamefont {Gaskill}}, \bibinfo
		{author} {\bibfnamefont {T.}~\bibnamefont {Reinecke}},\ and\ \bibinfo
		{author} {\bibfnamefont {S.~G.}\ \bibnamefont {Carter}},\ }\bibfield  {title}
	{\bibinfo {title} {Resonant optical spin initialization and readout of single
			silicon vacancies in $4H$-$\mathrm{Si}\mathrm{C}$},\ }\href
	{https://doi.org/10.1103/PhysRevApplied.11.024013} {\bibfield  {journal}
		{\bibinfo  {journal} {Phys. Rev. Applied}\ }\textbf {\bibinfo {volume}
			{11}},\ \bibinfo {pages} {024013} (\bibinfo {year} {2019})}\BibitemShut
	{NoStop}%
	\bibitem [{\citenamefont {Siyushev}\ \emph {et~al.}(2010)\citenamefont
		{Siyushev}, \citenamefont {Kaiser}, \citenamefont {Jacques}, \citenamefont
		{Gerhardt}, \citenamefont {Bischof}, \citenamefont {Fedder}, \citenamefont
		{Dodson}, \citenamefont {Markham}, \citenamefont {Twitchen}, \citenamefont
		{Jelezko},\ and\ \citenamefont {Wrachtrup}}]{Siyushev2010}%
	\BibitemOpen
	\bibfield  {author} {\bibinfo {author} {\bibfnamefont {P.}~\bibnamefont
			{Siyushev}}, \bibinfo {author} {\bibfnamefont {F.}~\bibnamefont {Kaiser}},
		\bibinfo {author} {\bibfnamefont {V.}~\bibnamefont {Jacques}}, \bibinfo
		{author} {\bibfnamefont {I.}~\bibnamefont {Gerhardt}}, \bibinfo {author}
		{\bibfnamefont {S.}~\bibnamefont {Bischof}}, \bibinfo {author} {\bibfnamefont
			{H.}~\bibnamefont {Fedder}}, \bibinfo {author} {\bibfnamefont
			{J.}~\bibnamefont {Dodson}}, \bibinfo {author} {\bibfnamefont
			{M.}~\bibnamefont {Markham}}, \bibinfo {author} {\bibfnamefont
			{D.}~\bibnamefont {Twitchen}}, \bibinfo {author} {\bibfnamefont
			{F.}~\bibnamefont {Jelezko}},\ and\ \bibinfo {author} {\bibfnamefont
			{J.}~\bibnamefont {Wrachtrup}},\ }\bibfield  {title} {\bibinfo {title}
		{Monolithic diamond optics for single photon detection},\ }\href
	{https://doi.org/10.1063/1.3519849} {\bibfield  {journal} {\bibinfo
			{journal} {Appl. Phys. Lett.}\ }\textbf {\bibinfo {volume} {97}},\ \bibinfo
		{pages} {241902} (\bibinfo {year} {2010})}\BibitemShut {NoStop}%
	\bibitem [{\citenamefont {Nagy}\ \emph {et~al.}(2018)\citenamefont {Nagy},
		\citenamefont {Widmann}, \citenamefont {Niethammer}, \citenamefont {Dasari},
		\citenamefont {Gerhardt}, \citenamefont {Soykal}, \citenamefont {Radulaski},
		\citenamefont {Ohshima}, \citenamefont {Vu\ifmmode \check{c}\else
			\v{c}\fi{}kovi\ifmmode~\acute{c}\else \'{c}\fi{}}, \citenamefont {Son},
		\citenamefont {Ivanov}, \citenamefont {Economou}, \citenamefont {Bonato},
		\citenamefont {Lee},\ and\ \citenamefont {Wrachtrup}}]{Nagy2018}%
	\BibitemOpen
	\bibfield  {author} {\bibinfo {author} {\bibfnamefont {R.}~\bibnamefont
			{Nagy}}, \bibinfo {author} {\bibfnamefont {M.}~\bibnamefont {Widmann}},
		\bibinfo {author} {\bibfnamefont {M.}~\bibnamefont {Niethammer}}, \bibinfo
		{author} {\bibfnamefont {D.~B.~R.}\ \bibnamefont {Dasari}}, \bibinfo {author}
		{\bibfnamefont {I.}~\bibnamefont {Gerhardt}}, \bibinfo {author}
		{\bibfnamefont {O.~O.}\ \bibnamefont {Soykal}}, \bibinfo {author}
		{\bibfnamefont {M.}~\bibnamefont {Radulaski}}, \bibinfo {author}
		{\bibfnamefont {T.}~\bibnamefont {Ohshima}}, \bibinfo {author} {\bibfnamefont
			{J.}~\bibnamefont {Vu\ifmmode \check{c}\else
				\v{c}\fi{}kovi\ifmmode~\acute{c}\else \'{c}\fi{}}}, \bibinfo {author}
		{\bibfnamefont {N.~T.}\ \bibnamefont {Son}}, \bibinfo {author} {\bibfnamefont
			{I.~G.}\ \bibnamefont {Ivanov}}, \bibinfo {author} {\bibfnamefont {S.~E.}\
			\bibnamefont {Economou}}, \bibinfo {author} {\bibfnamefont {C.}~\bibnamefont
			{Bonato}}, \bibinfo {author} {\bibfnamefont {S.-Y.}\ \bibnamefont {Lee}},\
		and\ \bibinfo {author} {\bibfnamefont {J.}~\bibnamefont {Wrachtrup}},\
	}\bibfield  {title} {\bibinfo {title} {Quantum properties of dichroic silicon
			vacancies in silicon carbide},\ }\href
	{https://doi.org/10.1103/PhysRevApplied.9.034022} {\bibfield  {journal}
		{\bibinfo  {journal} {Phys. Rev. Applied}\ }\textbf {\bibinfo {volume} {9}},\
		\bibinfo {pages} {034022} (\bibinfo {year} {2018})}\BibitemShut {NoStop}%
	\bibitem [{\citenamefont {Gotardo}\ \emph {et~al.}(2017)\citenamefont
		{Gotardo}, \citenamefont {Cocca}, \citenamefont {Acunha}, \citenamefont
		{Longoni}, \citenamefont {Toldo}, \citenamefont {Gonçalves}, \citenamefont
		{Iglesias},\ and\ \citenamefont {{De Boni}}}]{GOTARDO2017}%
	\BibitemOpen
	\bibfield  {author} {\bibinfo {author} {\bibfnamefont {F.}~\bibnamefont
			{Gotardo}}, \bibinfo {author} {\bibfnamefont {L.~H.}\ \bibnamefont {Cocca}},
		\bibinfo {author} {\bibfnamefont {T.~V.}\ \bibnamefont {Acunha}}, \bibinfo
		{author} {\bibfnamefont {A.}~\bibnamefont {Longoni}}, \bibinfo {author}
		{\bibfnamefont {J.}~\bibnamefont {Toldo}}, \bibinfo {author} {\bibfnamefont
			{P.~F.}\ \bibnamefont {Gonçalves}}, \bibinfo {author} {\bibfnamefont
			{B.~A.}\ \bibnamefont {Iglesias}},\ and\ \bibinfo {author} {\bibfnamefont
			{L.}~\bibnamefont {{De Boni}}},\ }\bibfield  {title} {\bibinfo {title}
		{Investigating the intersystem crossing rate and triplet quantum yield of
			protoporphyrin ix by means of pulse train fluorescence technique},\ }\href
	{https://doi.org/https://doi.org/10.1016/j.cplett.2017.02.055} {\bibfield
		{journal} {\bibinfo  {journal} {Chemical Physics Letters}\ }\textbf {\bibinfo
			{volume} {674}},\ \bibinfo {pages} {48} (\bibinfo {year} {2017})}\BibitemShut
	{NoStop}%
	\bibitem [{\citenamefont {Niethammer}\ \emph {et~al.}(2019)\citenamefont
		{Niethammer}, \citenamefont {Widmann}, \citenamefont {Rendler}, \citenamefont
		{Morioka}, \citenamefont {Chen}, \citenamefont {St{\"o}hr}, \citenamefont
		{Hassan}, \citenamefont {Onoda}, \citenamefont {Ohshima}, \citenamefont
		{Lee}, \citenamefont {Mukherjee}, \citenamefont {Isoya}, \citenamefont
		{Son},\ and\ \citenamefont {Wrachtrup}}]{Niethammer2019}%
	\BibitemOpen
	\bibfield  {author} {\bibinfo {author} {\bibfnamefont {M.}~\bibnamefont
			{Niethammer}}, \bibinfo {author} {\bibfnamefont {M.}~\bibnamefont {Widmann}},
		\bibinfo {author} {\bibfnamefont {T.}~\bibnamefont {Rendler}}, \bibinfo
		{author} {\bibfnamefont {N.}~\bibnamefont {Morioka}}, \bibinfo {author}
		{\bibfnamefont {Y.-C.}\ \bibnamefont {Chen}}, \bibinfo {author}
		{\bibfnamefont {R.}~\bibnamefont {St{\"o}hr}}, \bibinfo {author}
		{\bibfnamefont {J.~U.}\ \bibnamefont {Hassan}}, \bibinfo {author}
		{\bibfnamefont {S.}~\bibnamefont {Onoda}}, \bibinfo {author} {\bibfnamefont
			{T.}~\bibnamefont {Ohshima}}, \bibinfo {author} {\bibfnamefont {S.-Y.}\
			\bibnamefont {Lee}}, \bibinfo {author} {\bibfnamefont {A.}~\bibnamefont
			{Mukherjee}}, \bibinfo {author} {\bibfnamefont {J.}~\bibnamefont {Isoya}},
		\bibinfo {author} {\bibfnamefont {N.~T.}\ \bibnamefont {Son}},\ and\ \bibinfo
		{author} {\bibfnamefont {J.}~\bibnamefont {Wrachtrup}},\ }\bibfield  {title}
	{\bibinfo {title} {Coherent electrical readout of defect spins in silicon
			carbide by photo-ionization at ambient conditions},\ }\href
	{https://doi.org/10.1038/s41467-019-13545-z} {\bibfield  {journal} {\bibinfo
			{journal} {Nature Communications}\ }\textbf {\bibinfo {volume} {10}},\
		\bibinfo {pages} {5569} (\bibinfo {year} {2019})}\BibitemShut {NoStop}%
	\bibitem [{\citenamefont {Robledo}\ \emph
		{et~al.}(2011{\natexlab{b}})\citenamefont {Robledo}, \citenamefont
		{Childress}, \citenamefont {Bernien}, \citenamefont {Hensen}, \citenamefont
		{Alkemade},\ and\ \citenamefont {Hanson}}]{Robledo2011}%
	\BibitemOpen
	\bibfield  {author} {\bibinfo {author} {\bibfnamefont {L.}~\bibnamefont
			{Robledo}}, \bibinfo {author} {\bibfnamefont {L.}~\bibnamefont {Childress}},
		\bibinfo {author} {\bibfnamefont {H.}~\bibnamefont {Bernien}}, \bibinfo
		{author} {\bibfnamefont {B.}~\bibnamefont {Hensen}}, \bibinfo {author}
		{\bibfnamefont {P.~F.~A.}\ \bibnamefont {Alkemade}},\ and\ \bibinfo {author}
		{\bibfnamefont {R.}~\bibnamefont {Hanson}},\ }\bibfield  {title} {\bibinfo
		{title} {High-fidelity projective read-out of a solid-state spin quantum
			register},\ }\href {https://doi.org/10.1038/nature10401} {\bibfield
		{journal} {\bibinfo  {journal} {Nature}\ }\textbf {\bibinfo {volume} {477}},\
		\bibinfo {pages} {574} (\bibinfo {year} {2011}{\natexlab{b}})}\BibitemShut
	{NoStop}%
	\bibitem [{\citenamefont {Sipahigil}\ \emph {et~al.}(2016)\citenamefont
		{Sipahigil}, \citenamefont {Evans}, \citenamefont {Sukachev}, \citenamefont
		{Burek}, \citenamefont {Borregaard}, \citenamefont {Bhaskar}, \citenamefont
		{Nguyen}, \citenamefont {Pacheco}, \citenamefont {Atikian}, \citenamefont
		{Meuwly}, \citenamefont {Camacho}, \citenamefont {Jelezko}, \citenamefont
		{Bielejec}, \citenamefont {Park}, \citenamefont {Lon{\v c}ar},\ and\
		\citenamefont {Lukin}}]{Sipahigil2016}%
	\BibitemOpen
	\bibfield  {author} {\bibinfo {author} {\bibfnamefont {A.}~\bibnamefont
			{Sipahigil}}, \bibinfo {author} {\bibfnamefont {R.~E.}\ \bibnamefont
			{Evans}}, \bibinfo {author} {\bibfnamefont {D.~D.}\ \bibnamefont {Sukachev}},
		\bibinfo {author} {\bibfnamefont {M.~J.}\ \bibnamefont {Burek}}, \bibinfo
		{author} {\bibfnamefont {J.}~\bibnamefont {Borregaard}}, \bibinfo {author}
		{\bibfnamefont {M.~K.}\ \bibnamefont {Bhaskar}}, \bibinfo {author}
		{\bibfnamefont {C.~T.}\ \bibnamefont {Nguyen}}, \bibinfo {author}
		{\bibfnamefont {J.~L.}\ \bibnamefont {Pacheco}}, \bibinfo {author}
		{\bibfnamefont {H.~A.}\ \bibnamefont {Atikian}}, \bibinfo {author}
		{\bibfnamefont {C.}~\bibnamefont {Meuwly}}, \bibinfo {author} {\bibfnamefont
			{R.~M.}\ \bibnamefont {Camacho}}, \bibinfo {author} {\bibfnamefont
			{F.}~\bibnamefont {Jelezko}}, \bibinfo {author} {\bibfnamefont
			{E.}~\bibnamefont {Bielejec}}, \bibinfo {author} {\bibfnamefont
			{H.}~\bibnamefont {Park}}, \bibinfo {author} {\bibfnamefont {M.}~\bibnamefont
			{Lon{\v c}ar}},\ and\ \bibinfo {author} {\bibfnamefont {M.~D.}\ \bibnamefont
			{Lukin}},\ }\bibfield  {title} {\bibinfo {title} {An integrated diamond
			nanophotonics platform for quantum-optical networks},\ }\href
	{https://doi.org/10.1126/science.aah6875} {\bibfield  {journal} {\bibinfo
			{journal} {Science}\ }\textbf {\bibinfo {volume} {354}},\ \bibinfo {pages}
		{847} (\bibinfo {year} {2016})}\BibitemShut {NoStop}%
	\bibitem [{\citenamefont {Bracher}\ \emph {et~al.}(2017)\citenamefont
		{Bracher}, \citenamefont {Zhang},\ and\ \citenamefont {Hu}}]{Bracher2017}%
	\BibitemOpen
	\bibfield  {author} {\bibinfo {author} {\bibfnamefont {D.~O.}\ \bibnamefont
			{Bracher}}, \bibinfo {author} {\bibfnamefont {X.}~\bibnamefont {Zhang}},\
		and\ \bibinfo {author} {\bibfnamefont {E.~L.}\ \bibnamefont {Hu}},\
	}\bibfield  {title} {\bibinfo {title} {Selective purcell enhancement of two
			closely linked zero-phonon transitions of a silicon carbide color center},\
	}\href {https://doi.org/10.1073/pnas.1704219114} {\bibfield  {journal}
		{\bibinfo  {journal} {PNAS}\ }\textbf {\bibinfo {volume} {114}},\ \bibinfo
		{pages} {4060} (\bibinfo {year} {2017})}\BibitemShut {NoStop}%
	\bibitem [{\citenamefont {Crook}\ \emph {et~al.}(2020)\citenamefont {Crook},
		\citenamefont {Anderson}, \citenamefont {Miao}, \citenamefont {Bourassa},
		\citenamefont {Lee}, \citenamefont {Bayliss}, \citenamefont {Bracher},
		\citenamefont {Zhang}, \citenamefont {Abe}, \citenamefont {Ohshima},
		\citenamefont {Hu},\ and\ \citenamefont {Awschalom}}]{Crook2020}%
	\BibitemOpen
	\bibfield  {author} {\bibinfo {author} {\bibfnamefont {A.~L.}\ \bibnamefont
			{Crook}}, \bibinfo {author} {\bibfnamefont {C.~P.}\ \bibnamefont {Anderson}},
		\bibinfo {author} {\bibfnamefont {K.~C.}\ \bibnamefont {Miao}}, \bibinfo
		{author} {\bibfnamefont {A.}~\bibnamefont {Bourassa}}, \bibinfo {author}
		{\bibfnamefont {H.}~\bibnamefont {Lee}}, \bibinfo {author} {\bibfnamefont
			{S.~L.}\ \bibnamefont {Bayliss}}, \bibinfo {author} {\bibfnamefont {D.~O.}\
			\bibnamefont {Bracher}}, \bibinfo {author} {\bibfnamefont {X.}~\bibnamefont
			{Zhang}}, \bibinfo {author} {\bibfnamefont {H.}~\bibnamefont {Abe}}, \bibinfo
		{author} {\bibfnamefont {T.}~\bibnamefont {Ohshima}}, \bibinfo {author}
		{\bibfnamefont {E.~L.}\ \bibnamefont {Hu}},\ and\ \bibinfo {author}
		{\bibfnamefont {D.~D.}\ \bibnamefont {Awschalom}},\ }\bibfield  {title}
	{\bibinfo {title} {Purcell enhancement of a single silicon carbide color
			center with coherent spin control},\ }\href
	{https://doi.org/10.1021/acs.nanolett.0c00339} {\bibfield  {journal}
		{\bibinfo  {journal} {Nano Lett.}\ }\textbf {\bibinfo {volume} {20}},\
		\bibinfo {pages} {3427} (\bibinfo {year} {2020})}\BibitemShut {NoStop}%
	\bibitem [{\citenamefont {Lukin}\ \emph
		{et~al.}(2020{\natexlab{b}})\citenamefont {Lukin}, \citenamefont {Dory},
		\citenamefont {Guidry}, \citenamefont {Yang}, \citenamefont {Mishra},
		\citenamefont {Trivedi}, \citenamefont {Radulaski}, \citenamefont {Sun},
		\citenamefont {Vercruysse}, \citenamefont {Ahn},\ and\ \citenamefont
		{Vu{\v{c}}kovi{\'{c}}}}]{Lukin2020}%
	\BibitemOpen
	\bibfield  {author} {\bibinfo {author} {\bibfnamefont {D.~M.}\ \bibnamefont
			{Lukin}}, \bibinfo {author} {\bibfnamefont {C.}~\bibnamefont {Dory}},
		\bibinfo {author} {\bibfnamefont {M.~A.}\ \bibnamefont {Guidry}}, \bibinfo
		{author} {\bibfnamefont {K.~Y.}\ \bibnamefont {Yang}}, \bibinfo {author}
		{\bibfnamefont {S.~D.}\ \bibnamefont {Mishra}}, \bibinfo {author}
		{\bibfnamefont {R.}~\bibnamefont {Trivedi}}, \bibinfo {author} {\bibfnamefont
			{M.}~\bibnamefont {Radulaski}}, \bibinfo {author} {\bibfnamefont
			{S.}~\bibnamefont {Sun}}, \bibinfo {author} {\bibfnamefont {D.}~\bibnamefont
			{Vercruysse}}, \bibinfo {author} {\bibfnamefont {G.~H.}\ \bibnamefont
			{Ahn}},\ and\ \bibinfo {author} {\bibfnamefont {J.}~\bibnamefont
			{Vu{\v{c}}kovi{\'{c}}}},\ }\bibfield  {title} {\bibinfo {title}
		{4h-silicon-carbide-on-insulator for integrated quantum and nonlinear
			photonics},\ }\href {https://doi.org/10.1038/s41566-019-0556-6} {\bibfield
		{journal} {\bibinfo  {journal} {Nat. Photonics}\ }\textbf {\bibinfo {volume}
			{14}},\ \bibinfo {pages} {330} (\bibinfo {year}
		{2020}{\natexlab{b}})}\BibitemShut {NoStop}%
	\bibitem [{\citenamefont {Mohtashami}\ and\ \citenamefont
		{Koenderink}(2013)}]{Mohtashami2013}%
	\BibitemOpen
	\bibfield  {author} {\bibinfo {author} {\bibfnamefont {A.}~\bibnamefont
			{Mohtashami}}\ and\ \bibinfo {author} {\bibfnamefont {A.~F.}\ \bibnamefont
			{Koenderink}},\ }\bibfield  {title} {\bibinfo {title} {Suitability of
			nanodiamond nitrogen{\textendash}vacancy centers for spontaneous emission
			control experiments},\ }\href {https://doi.org/10.1088/1367-2630/15/4/043017}
	{\bibfield  {journal} {\bibinfo  {journal} {New J. Phys.}\ }\textbf {\bibinfo
			{volume} {15}},\ \bibinfo {pages} {043017} (\bibinfo {year}
		{2013})}\BibitemShut {NoStop}%
	\bibitem [{\citenamefont {Inam}\ \emph {et~al.}(2014)\citenamefont {Inam},
		\citenamefont {Steel},\ and\ \citenamefont {Castelletto}}]{INAM2014}%
	\BibitemOpen
	\bibfield  {author} {\bibinfo {author} {\bibfnamefont {F.}~\bibnamefont
			{Inam}}, \bibinfo {author} {\bibfnamefont {M.}~\bibnamefont {Steel}},\ and\
		\bibinfo {author} {\bibfnamefont {S.}~\bibnamefont {Castelletto}},\
	}\bibfield  {title} {\bibinfo {title} {Effects of the hosting
			nano-environment modifications on nv centres fluorescence emission},\ }\href
	{https://doi.org/https://doi.org/10.1016/j.diamond.2014.03.008} {\bibfield
		{journal} {\bibinfo  {journal} {Diam. Relat. Mater.}\ }\textbf {\bibinfo
			{volume} {45}},\ \bibinfo {pages} {64} (\bibinfo {year} {2014})}\BibitemShut
	{NoStop}%
	\bibitem [{\citenamefont {Alkauskas}\ \emph {et~al.}(2014)\citenamefont
		{Alkauskas}, \citenamefont {Buckley}, \citenamefont {Awschalom},\ and\
		\citenamefont {de~Walle}}]{Alkauskas_2014}%
	\BibitemOpen
	\bibfield  {author} {\bibinfo {author} {\bibfnamefont {A.}~\bibnamefont
			{Alkauskas}}, \bibinfo {author} {\bibfnamefont {B.~B.}\ \bibnamefont
			{Buckley}}, \bibinfo {author} {\bibfnamefont {D.~D.}\ \bibnamefont
			{Awschalom}},\ and\ \bibinfo {author} {\bibfnamefont {C.~G.~V.}\ \bibnamefont
			{de~Walle}},\ }\bibfield  {title} {\bibinfo {title} {First-principles theory
			of the luminescence lineshape for the triplet transition in diamond {NV}
			centres},\ }\href {https://doi.org/10.1088/1367-2630/16/7/073026} {\bibfield
		{journal} {\bibinfo  {journal} {New J. Phys.}\ }\textbf {\bibinfo {volume}
			{16}},\ \bibinfo {pages} {073026} (\bibinfo {year} {2014})}\BibitemShut
	{NoStop}%
	\bibitem [{\citenamefont {Stoneham}(2001)}]{Stoneham2007}%
	\BibitemOpen
	\bibfield  {author} {\bibinfo {author} {\bibfnamefont {A.~M.}\ \bibnamefont
			{Stoneham}},\ }\href@noop {} {\emph {\bibinfo {title} {Theory of Defects in
				Solids: Electronic Structure of Defects in Insulators and Semiconductors}}}\
	(\bibinfo  {publisher} {Oxford University Press},\ \bibinfo {address}
	{Ocford},\ \bibinfo {year} {2001})\BibitemShut {NoStop}%
	\bibitem [{\citenamefont {Fischer}\ \emph {et~al.}(2017)\citenamefont
		{Fischer}, \citenamefont {Hanschke}, \citenamefont {Kremser}, \citenamefont
		{Finley}, \citenamefont {Müller},\ and\ \citenamefont
		{Vu{\v{c}}kovi{\'{c}}}}]{Fischer2017}%
	\BibitemOpen
	\bibfield  {author} {\bibinfo {author} {\bibfnamefont {K.~A.}\ \bibnamefont
			{Fischer}}, \bibinfo {author} {\bibfnamefont {L.}~\bibnamefont {Hanschke}},
		\bibinfo {author} {\bibfnamefont {M.}~\bibnamefont {Kremser}}, \bibinfo
		{author} {\bibfnamefont {J.~J.}\ \bibnamefont {Finley}}, \bibinfo {author}
		{\bibfnamefont {K.}~\bibnamefont {Müller}},\ and\ \bibinfo {author}
		{\bibfnamefont {J.}~\bibnamefont {Vu{\v{c}}kovi{\'{c}}}},\ }\bibfield
	{title} {\bibinfo {title} {Pulsed rabi oscillations in quantum two-level
			systems: beyond the area theorem},\ }\href
	{https://doi.org/10.1088/2058-9565/aa9269} {\bibfield  {journal} {\bibinfo
			{journal} {Quantum Sci. Technol.}\ }\textbf {\bibinfo {volume} {3}},\
		\bibinfo {pages} {014006} (\bibinfo {year} {2017})}\BibitemShut {NoStop}%
	\bibitem [{\citenamefont {Becker}\ and\ \citenamefont
		{Becher}(2017)}]{Becher2017}%
	\BibitemOpen
	\bibfield  {author} {\bibinfo {author} {\bibfnamefont {J.~N.}\ \bibnamefont
			{Becker}}\ and\ \bibinfo {author} {\bibfnamefont {C.}~\bibnamefont
			{Becher}},\ }\bibfield  {title} {\bibinfo {title} {Coherence properties and
			quantum control of silicon vacancy color centers in diamond},\ }\href
	{https://doi.org/https://doi.org/10.1002/pssa.201700586} {\bibfield
		{journal} {\bibinfo  {journal} {Phys. Status Solidi A}\ }\textbf {\bibinfo
			{volume} {214}},\ \bibinfo {pages} {1700586} (\bibinfo {year}
		{2017})}\BibitemShut {NoStop}%
	\bibitem [{\citenamefont {Riedel}\ \emph {et~al.}(2017)\citenamefont {Riedel},
		\citenamefont {S\"ollner}, \citenamefont {Shields}, \citenamefont
		{Starosielec}, \citenamefont {Appel}, \citenamefont {Neu}, \citenamefont
		{Maletinsky},\ and\ \citenamefont {Warburton}}]{Riedel2017}%
	\BibitemOpen
	\bibfield  {author} {\bibinfo {author} {\bibfnamefont {D.}~\bibnamefont
			{Riedel}}, \bibinfo {author} {\bibfnamefont {I.}~\bibnamefont {S\"ollner}},
		\bibinfo {author} {\bibfnamefont {B.~J.}\ \bibnamefont {Shields}}, \bibinfo
		{author} {\bibfnamefont {S.}~\bibnamefont {Starosielec}}, \bibinfo {author}
		{\bibfnamefont {P.}~\bibnamefont {Appel}}, \bibinfo {author} {\bibfnamefont
			{E.}~\bibnamefont {Neu}}, \bibinfo {author} {\bibfnamefont {P.}~\bibnamefont
			{Maletinsky}},\ and\ \bibinfo {author} {\bibfnamefont {R.~J.}\ \bibnamefont
			{Warburton}},\ }\bibfield  {title} {\bibinfo {title} {Deterministic
			enhancement of coherent photon generation from a nitrogen-vacancy center in
			ultrapure diamond},\ }\href {https://doi.org/10.1103/PhysRevX.7.031040}
	{\bibfield  {journal} {\bibinfo  {journal} {Phys. Rev. X}\ }\textbf {\bibinfo
			{volume} {7}},\ \bibinfo {pages} {031040} (\bibinfo {year}
		{2017})}\BibitemShut {NoStop}%
	\bibitem [{\citenamefont {Soykal}\ \emph {et~al.}(2016)\citenamefont {Soykal},
		\citenamefont {Dev},\ and\ \citenamefont {Economou}}]{Soykal2016}%
	\BibitemOpen
	\bibfield  {author} {\bibinfo {author} {\bibfnamefont {{\"O}.~O.}\
			\bibnamefont {Soykal}}, \bibinfo {author} {\bibfnamefont {P.}~\bibnamefont
			{Dev}},\ and\ \bibinfo {author} {\bibfnamefont {S.~E.}\ \bibnamefont
			{Economou}},\ }\bibfield  {title} {\bibinfo {title} {Silicon vacancy center
			in $4H$-SiC: Electronic structure and spin-photon interfaces},\ }\href
	{https://doi.org/10.1103/PhysRevB.93.081207} {\bibfield  {journal} {\bibinfo
			{journal} {Phys. Rev. B}\ }\textbf {\bibinfo {volume} {93}},\ \bibinfo
		{pages} {081207} (\bibinfo {year} {2016})}\BibitemShut {NoStop}%
	\bibitem [{\citenamefont {Shang}\ \emph {et~al.}(2020)\citenamefont {Shang},
		\citenamefont {Hashemi}, \citenamefont {Berenc\'en}, \citenamefont {Komsa},
		\citenamefont {Erhart}, \citenamefont {Zhou}, \citenamefont {Helm},
		\citenamefont {Krasheninnikov},\ and\ \citenamefont {Astakhov}}]{Shang2020}%
	\BibitemOpen
	\bibfield  {author} {\bibinfo {author} {\bibfnamefont {Z.}~\bibnamefont
			{Shang}}, \bibinfo {author} {\bibfnamefont {A.}~\bibnamefont {Hashemi}},
		\bibinfo {author} {\bibfnamefont {Y.}~\bibnamefont {Berenc\'en}}, \bibinfo
		{author} {\bibfnamefont {H.-P.}\ \bibnamefont {Komsa}}, \bibinfo {author}
		{\bibfnamefont {P.}~\bibnamefont {Erhart}}, \bibinfo {author} {\bibfnamefont
			{S.}~\bibnamefont {Zhou}}, \bibinfo {author} {\bibfnamefont {M.}~\bibnamefont
			{Helm}}, \bibinfo {author} {\bibfnamefont {A.~V.}\ \bibnamefont
			{Krasheninnikov}},\ and\ \bibinfo {author} {\bibfnamefont {G.~V.}\
			\bibnamefont {Astakhov}},\ }\bibfield  {title} {\bibinfo {title} {Local
			vibrational modes of si vacancy spin qubits in sic},\ }\href
	{https://doi.org/10.1103/PhysRevB.101.144109} {\bibfield  {journal} {\bibinfo
			{journal} {Phys. Rev. B}\ }\textbf {\bibinfo {volume} {101}},\ \bibinfo
		{pages} {144109} (\bibinfo {year} {2020})}\BibitemShut {NoStop}%
	\bibitem [{\citenamefont {Burshtein}(2010)}]{Burstein2010}%
	\BibitemOpen
	\bibfield  {author} {\bibinfo {author} {\bibfnamefont {Z.}~\bibnamefont
			{Burshtein}},\ }\bibfield  {title} {\bibinfo {title} {{Radiative,
				nonradiative, and mixed-decay transitions of rare-earth ions in dielectric
				media}},\ }\href {https://doi.org/10.1117/1.3483907} {\bibfield  {journal}
		{\bibinfo  {journal} {Optical Engineering}\ }\textbf {\bibinfo {volume}
			{49}},\ \bibinfo {pages} {1 } (\bibinfo {year} {2010})}\BibitemShut {NoStop}%
	\bibitem [{\citenamefont {Larkins}\ and\ \citenamefont
		{Stoneham}(1971)}]{Larkins1971}%
	\BibitemOpen
	\bibfield  {author} {\bibinfo {author} {\bibfnamefont {F.~P.}\ \bibnamefont
			{Larkins}}\ and\ \bibinfo {author} {\bibfnamefont {A.~M.}\ \bibnamefont
			{Stoneham}},\ }\bibfield  {title} {\bibinfo {title} {Lattice distortion near
			vacancies in diamond and silicon. ii},\ }\href
	{https://doi.org/10.1088/0022-3719/4/2/003} {\bibfield  {journal} {\bibinfo
			{journal} {J. Phys. C: Solid St. Phys.}\ }\textbf {\bibinfo {volume} {4}},\
		\bibinfo {pages} {154} (\bibinfo {year} {1971})}\BibitemShut {NoStop}%
	\bibitem [{\citenamefont {Bockstedte}\ \emph {et~al.}(2003)\citenamefont
		{Bockstedte}, \citenamefont {Mattausch},\ and\ \citenamefont
		{Pankratov}}]{Bockstedte2003}%
	\BibitemOpen
	\bibfield  {author} {\bibinfo {author} {\bibfnamefont {M.}~\bibnamefont
			{Bockstedte}}, \bibinfo {author} {\bibfnamefont {A.}~\bibnamefont
			{Mattausch}},\ and\ \bibinfo {author} {\bibfnamefont {O.}~\bibnamefont
			{Pankratov}},\ }\bibfield  {title} {\bibinfo {title} {Ab initio study of the
			migration of intrinsic defects in $3c\ensuremath{-}\mathrm{SiC}$},\ }\href
	{https://doi.org/10.1103/PhysRevB.68.205201} {\bibfield  {journal} {\bibinfo
			{journal} {Phys. Rev. B}\ }\textbf {\bibinfo {volume} {68}},\ \bibinfo
		{pages} {205201} (\bibinfo {year} {2003})}\BibitemShut {NoStop}%
	\bibitem [{\citenamefont {Chapman}\ and\ \citenamefont
		{Plakhotnik}(2011)}]{CHAPMAN2011}%
	\BibitemOpen
	\bibfield  {author} {\bibinfo {author} {\bibfnamefont {R.}~\bibnamefont
			{Chapman}}\ and\ \bibinfo {author} {\bibfnamefont {T.}~\bibnamefont
			{Plakhotnik}},\ }\bibfield  {title} {\bibinfo {title} {Quantitative
			luminescence microscopy on nitrogen-vacancy centres in diamond: Saturation
			effects under pulsed excitation},\ }\href
	{https://doi.org/https://doi.org/10.1016/j.cplett.2011.03.057} {\bibfield
		{journal} {\bibinfo  {journal} {Chemical Physics Letters}\ }\textbf {\bibinfo
			{volume} {507}},\ \bibinfo {pages} {190} (\bibinfo {year}
		{2011})}\BibitemShut {NoStop}%
\end{thebibliography}

%

\end{document}